\documentstyle[floats,prd,aps,epsf]{revtex}
\newcommand{\beq}{\begin{equation}}
\newcommand{\eeq}{\end{equation}}
\newcommand{\beqn}{\begin{eqnarray}}
\newcommand{\eeqn}{\end{eqnarray}}

\newcommand{\cB}{{\cal{B}}}

\newcommand{\pa}{\partial}

\newcommand{\Ptot}{P_{\rm tot}}
\newcommand{\varep}{\varepsilon}

\def\agt{\mathrel{\raise.3ex\hbox{$>$}\mkern-14mu\lower0.6ex\hbox{$\sim$}}}
\def\alt{\mathrel{\raise.3ex\hbox{$<$}\mkern-14mu\lower0.6ex\hbox{$\sim$}}}

\draft
\begin{document}

\begin{center}
{\large\bf{Magnetohydrodynamics in full general relativity: 
Formulation and tests
}}
~~\\
~~\\
Masaru Shibata and Yu-ichiou Sekiguchi
~~\\
~~\\
{\em Graduate School of Arts and Sciences, 
University of Tokyo, Komaba, Meguro, Tokyo 153-8902, Japan 
}
\end{center}

\begin{abstract}
A new implementation for magnetohydrodynamics (MHD) simulations in
full general relativity (involving dynamical spacetimes) is presented. 
In our implementation, Einstein's evolution equations are evolved by a
BSSN formalism, MHD equations by a high-resolution central scheme, and
induction equation by a constraint transport method. We perform 
numerical simulations for standard test problems in relativistic MHD,
including special relativistic magnetized shocks, general relativistic
magnetized Bondi flow in stationary spacetime, and a longterm
evolution for self-gravitating system composed of a neutron star and a
magnetized disk in full general relativity. In the final test, we
illustrate that our implementation can follow winding-up of the
magnetic field lines of magnetized and differentially rotating
accretion disks around a compact object until saturation, after which
magnetically driven wind and angular momentum transport inside the
disk turn on.
\end{abstract}

\section{Introduction}

Hydrodynamics simulation in general relativity is probably the best
theoretical approach for investigating dynamical phenomena in
relativistic astrophysics such as stellar core collapse to a neutron
star and a black hole, and the merger of binary neutron stars. 
In the past several years, this field has been extensively
developed (e.g., \cite{shiba99,font02,ref1,ref2,ref3,STU,STU2}) and,
as a result, now it is feasible to perform accurate
simulations of such general relativistic phenomena
for yielding scientific results (e.g., \cite{STU,STU2,SS3d,SS05} for 
our latest results). For example, with the current 
implementation, radiation reaction of gravitational waves in the merger
of binary neutron stars can be taken into account within $\sim 1\%$ error 
in an appropriate computational setting \cite{STU,STU2}.  This fact
illustrates that the numerical relativity is a robust approach 
for detailed theoretical study of astrophysical phenomena 
and gravitational waves emitted. 

However, so far, most of the scientific simulations in full general
relativity have been performed without taking into account detailed
effects except for general relativistic gravity and pure
hydrodynamics. For example, simplified ideal equations of state have
been adopted instead of realistic ones (but see \cite{STU2}).  Also,
the effect of magnetic fields has been neglected although it could
often play an important role in the astrophysical phenomena (but see
\cite{DLSS}). In the next stage of numerical relativity, it is
necessary to incorporate these effects for more realistic
simulations. As a step toward a more realistic simulation, we have
incorporated an implementation for ideal magnetohydrodynamics (MHD)
equations in fully general relativistic manner. In this paper, we
describe our approach for these equations and then present
numerical results for test problems computed by the new implementation.

Magnetic fields indeed play an important role in determining the
evolution of a number of relativistic objects.  In the astrophysical
context, the plasma is usually highly conducting, and hence, the
magnetic fields are frozen in the matter. This implies that a small
seed field can wind up and grow in the complex motion of the matter,
resulting in a significant effect in the dynamics of the matter such
as magnetically driven wind or jet and angular momentum
redistribution. Specifically, in the context of the general
relativistic astrophysics, the magnetic fields will play a role in the
following phenomena and objects: Stellar core collapse of 
magnetized massive stars to a protoneutron star \cite{AHM} or a black hole, 
stability of accretion disks (which are either non-self-gravitating
or self-gravitating) around black holes and neutron stars,
magnetic braking of differentially rotating neutron stars \cite{DLSS}
which are formed after merger of binary neutron stars \cite{STU,STU2}
and stellar core collapse \cite{Zweg,HD,Ott,SS3d,SS05}, and magnetically 
induced jet around the compact objects (e.g., \cite{koide}).
To clarify these phenomena, fully general relativistic 
MHD (GRMHD) simulation (involving dynamical spacetimes)
is probably the best theoretical approach.

In the past decade, numerical implementations for 
GRMHD simulation in the fixed gravitational field have been
extensively developed (e.g.,
\cite{MHD0,koide,komissarov2,gammie,VH,MHD2,HVH,Aloy}). In particular,
it is worth to mention that Refs. 
\cite{komissarov2,gammie,VH,MHD2,HVH} have recently presented 
implementations for which detailed tests have been carried out 
for confirmation of the reliability of their computation, 
in contrast with the attitude in an early work \cite{koide}.
They are applied for simulating magnetorotational instability (MRI) of
accretion disks and subsequently induced winds and jets
around black holes and neutron 
stars. On the other hand, little effort has been paid to numerical
implementations of fully GRMHD (in the dynamical
gravitational field). About 30 years ago, Wilson performed a simulation
for collapse of a magnetized star in the presence of poloidal
magnetic fields in general relativity. However, he assumes that the
three-metric is {\em conformally flat} \cite{wilson}, and hence, the
simulation is not fully general relativistic, although recent works
have indicated that the conformally flat approximation works well in the
axisymmetric collapse (e.g., compare results among \cite{HD}, 
\cite{SS}, and \cite{CFC}).
The first fully GRMHD simulation for stellar 
collapse was performed by Nakamura about 20 years ago \cite{NOK}. He
simulated collapse of nonrotating stars with poloidal magnetic fields
to investigate the criteria for formation of black holes and naked
singularities. Very recently, Duez et al. have presented a new implementation
capable of evolution for the Einstein-Maxwell-MHD equations for
general cases \cite{DLSS}. They report successful results for test
simulations. Valencia group has also developed a GRMHD implementation
very recently \cite{valencia}. 

In this paper, we present our new implementation for fully 
GRMHD which is similar to but in part different from that in \cite{DLSS}
\footnote{For instance, our formulation for Einstein's evolution equations,
gauge conditions, and 
our numerical scheme for GRMHD equations are different from those
in \cite{DLSS} as mentioned in Secs. II and III.}. 
As a first step toward scientific 
simulations, we have performed simulations in standard test problems
including special relativistic magnetized shocks, general
relativistic Bondi flow in stationary spacetime, and long term
evolution of fully general relativistic stars with magnetic fields.
We here report the successful results for these test problems. 

Before proceeding, we emphasize that it is important to
develop new GRMHD implementations. 
In the presence of magnetic fields, matter motion often becomes
turbulence-like due to MRIs in which a small scale structure
often grows most effectively \cite{BaHa}. 
Furthermore in the presence of general relativistic self-gravity
which has a nonlinear nature, the matter motion may be even
complicated.  Perhaps, the outputs from the simulations will contain
results which have not been well understood yet, and thus, 
are rich in new physics. Obviously high accuracy is 
required for such frontier simulation to confirm novel numerical results.
However, because of the restriction of computational resources, it is
often very difficult to get a well-resolved and completely 
convergent numerical result in fully general relativistic simulation. 
In such case, comparison among various results obtained by different
numerical implementations is crucial for checking the reliability of
the numerical results. From this point of view, it is important to
develop several numerical implementations in the community of
numerical relativity. By comparing several results computed by
different implementations,  
reliability of the numerical results will be improved each other. Our
implementation presented here will be useful not only for finding new
physics but also for checking numerical results by other
implementations such as that very recently presented
in \cite{DLSS,valencia}.

In Sec. II, we present formulations for Einstein, Maxwell, and GRMHD
equations. In Sec. III, numerical methods for solving GRMHD equations
are described. In Sec. IV, methods for a solution of initial value
problem in general relativity is presented. In Secs. V and VI,
numerical results for special and general relativistic test
simulations are shown.  In the final subsection of Sec. VI, we
illustrate that our implementation can follow growth of magnetic
fields of accretion disks in fully general relativistic simulation.
Sec. VII is devoted to a summary and a discussion. Throughout this
paper, we adopt the geometrical units in which $G=c=1$ where $G$ and
$c$ are the gravitational constant and the speed of light.  Latin and
Greek indices denote spatial components and spacetime components,
respectively. $\eta_{\mu\nu}$ and $\delta_{ij}(=\delta^{ij})$ denote
the flat spacetime metric and the Kronecker delta, respectively.

\section{Basic equations}

\subsection{Definition of variables} 

Basic equations consist of the Einstein equations, general relativistic
hydrodynamic equations, and Maxwell equations. In this
subsection, we define the variables used in these equations. 
The fundamental variables for geometry are 
$\alpha$: lapse function, $\beta^k$: shift vector, 
$\gamma_{ij}$: metric in three-dimensional spatial hypersurface,
and $K_{ij}$: extrinsic curvature. 
The spacetime metric $g_{\mu\nu}$ is written as
\beqn
g_{\mu\nu}=\gamma_{\mu\nu}-n_{\mu} n_{\nu}, 
\eeqn
where $n^{\mu}$ is a unit normal to a spacelike spatial hypersurface
$\Sigma$ and is written as
\beqn
n^{\mu}=\biggl({1 \over \alpha},
-{\beta^i \over \alpha}\biggr),~~~{\rm or}~~~
n_{\mu}=(-\alpha, 0). 
\eeqn

In the BSSN formalism \cite{BSSN}, one defines 
$\gamma\equiv \eta e^{12\phi}={\rm det}(\gamma_{ij})$: determinant
of $\gamma_{ij}$, 
$\tilde \gamma_{ij}=e^{-4\phi}\gamma_{ij}$: conformal three-metric, 
$K=K_k^{~k}$: trace of the extrinsic curvature, and
$\tilde A_{ij}\equiv e^{-4\phi}(K_{ij}-K\gamma_{ij}/3)$: 
a tracefree part of the extrinsic curvature. 
Here, $\eta$ denotes the determinant 
of flat metric; in the Cartesian coordinates, $\eta=1$, and in the
cylindrical coordinates $(\varpi, \varphi, z)$, $\eta=\varpi^2$. 
In the following, 
$\nabla_{\mu}$, $D_i$, and $\tilde D_i$ denote the covariant 
derivatives with respect to $g_{\mu\nu}$, $\gamma_{ij}$, and 
$\tilde \gamma_{ij}$, respectively. $\Delta$ and $\tilde \Delta$
denote the Laplacians with respect to $\gamma_{ij}$ and 
$\tilde \gamma_{ij}$. $R_{ij}$ and $\tilde R_{ij}$ denote the Ricci tensors 
with respect to $\gamma_{ij}$ and $\tilde \gamma_{ij}$, respectively. 

The fundamental variables in hydrodynamics are 
$\rho$: rest-mass density, 
$\varep$ : specific internal energy, 
$P$ : pressure, and $u^{\mu}$ : four velocity. From these
variables, we define the following
variables which often appear in the basic equations: 
\beqn
&& \rho_* \equiv \rho w e^{6\phi},\\
&&v^i \equiv
{dx^i \over dt}={u^i \over u^t}=-\beta^i + \gamma^{ij}{u_j \over u^t},\\
&&h \equiv 1+\varep + {P \over \rho},\\
&&w \equiv \alpha u^t. 
\eeqn
Here, $\rho_*$ is a weighted baryon rest mass density from which
the conserved baryon rest mass can be computed as 
\beqn
M_*= \int \rho_* \eta^{1/2} d^3x. 
\eeqn

The fundamental variable in the ideal MHD is only $b^{\mu}$: magnetic field.
The electric field $E^{\mu}$ in the comoving frame $F^{\mu\nu}u_{\nu}$
is assumed to be zero, and electric current $j^{\mu}$ is not 
explicitly necessary for evolving the field variables. 
Using the electromagnetic tensor $F^{\mu\nu}$, $b_{\mu}$
is defined by \cite{MTW}
\beqn
b_{\mu} \equiv -{1 \over 2}
\epsilon_{\mu\nu\alpha\beta}u^{\nu}F^{\alpha\beta}, \label{defb}
\eeqn
where $\epsilon_{\mu\nu\alpha\beta}$ is the Levi-Civita tensor
with $\epsilon_{t123}=\sqrt{-g}$ and $\epsilon^{t123}=-1/\sqrt{-g}$.
Equation (\ref{defb}) implies 
\beqn
b^{\mu}u_{\mu}=0. \label{eqbu}
\eeqn
Using Eq. (\ref{defb}), $F^{\mu\nu}$ in the ideal MHD is written as
\beqn
F^{\mu\nu}=\epsilon^{\mu\nu\alpha\beta}u_{\alpha}b_{\beta}, \label{eqFF}
\eeqn
and thus, it satisfies the ideal MHD condition
\beqn
F_{\mu\nu}u^{\nu}=0.
\eeqn
The dual tensor of $F_{\mu\nu}$ is defined by 
\beqn
F^*_{\mu\nu} \equiv {1 \over 2}\epsilon_{\mu\nu\alpha\beta} F^{\alpha\beta}
=b_{\mu} u_{\nu}- b_{\nu} u_{\mu}. 
\eeqn
For rewriting the induction equation for the magnetic fields
into a simple form (see Sec. II D), we define the three-magnetic field as
\beqn
\cB^i \equiv - e^{6\phi} \gamma^{i}_{~j} F^{*j\mu} n_{\mu}
=e^{6\phi} (w b^i - \alpha b^t u^i). \label{defB}
\eeqn
Here, we note that $\cB^t=0$ (i.e., $\cB^{\mu} n_{\mu}=0$),
and thus, $\cB_i=\gamma_{ij} \cB^j$. 
Equations (\ref{defB}) and (\ref{eqbu}) lead to 
\beqn
b^t = {\cB^{\mu} u_{\mu} \over \alpha e^{6\phi}}~~{\rm and}~~
b_i={1 \over w e^{6\phi}}\Big(\cB_i + \cB^j u_j u_i \Big)
\eeqn

Using the hydrodynamic and electromagnetic variables,
energy-momentum tensor is written as
\beqn
T_{\mu\nu}&=&T_{\mu\nu}^{\rm Fluid} + T_{\mu\nu}^{\rm EM}. 
\eeqn
$T_{\mu\nu}^{\rm Fluid}$ and $T_{\mu\nu}^{\rm EM}$ denote the
fluid and electromagnetic parts defined by 
\beqn
&&T_{\mu\nu}^{\rm Fluid}=
(\rho + \rho \varep + P)u_{\mu} u_{\nu} + P g_{\mu\nu}=
\rho h u_{\mu} u_{\nu} + P g_{\mu\nu}, \\
&&T_{\mu\nu}^{\rm EM}=
F_{\mu\sigma} F^{~\sigma}_{\nu}-{1 \over 4}g_{\mu\nu}
F_{\alpha\beta} F^{\alpha\beta} 
=\biggl({1 \over 2}g_{\mu\nu}+u_{\mu}u_{\nu}\biggr)b^2 -b_{\mu}b_{\nu},
\eeqn
where
\beqn
b^2=b_{\mu} b^{\mu}={\cB^2 + (\cB^i u_i)^2 \over w^2 e^{12\phi}}.
\eeqn
Thus, $T_{\mu\nu}$ is written as
\beqn
T_{\mu\nu}=(\rho h + b^2) u_{\mu} u_{\nu}+\Big(P + {1 \over 2}b^2\Big)
g_{\mu\nu}-b_{\mu}b_{\nu}. 
\eeqn
For the following, we define magnetic pressure and 
total pressure as $P_{\rm mag}=b^2/2$ and $\Ptot \equiv P+b^2/2$,
respectively. 

The (3+1) decomposition of $T_{\mu\nu}$ is
\beqn
&& \rho_{\rm H} \equiv T_{\mu\nu} n^{\mu}n^{\nu}=(\rho h +b^2) w^2
- \Ptot- (\alpha b^t)^2, \\
&& J_i \equiv  - T_{\mu\nu} n^{\mu}\gamma^{\nu}_{~i}
=(\rho h +b^2) w u_i -\alpha b^t b_i, \\
&& S_{ij} \equiv T_{\mu\nu} \gamma^{\mu}_{~i}\gamma^{\nu}_{~j}
=(\rho h + b^2) u_i u_j + \Ptot \gamma_{ij}- b_i b_j. 
\eeqn
Using these, the energy-momentum tensor is rewritten in the
form
\beqn
T_{\mu\nu}=\rho_{\rm H} n_{\mu} n_{\nu}
+ J_i \gamma^i_{~\mu} n_{\nu}+ J_i \gamma^i_{~\nu} n_{\mu}
+S_{ij} \gamma^i_{~\mu}\gamma^j_{~\nu}. \label{31t}
\eeqn
This form of the energy-momentum tensor is useful for deriving the
basic equations for GRMHD presented in Sec. II C.
For the following, we define
\beqn
S_0 \equiv e^{6\phi} \rho_{\rm H},\\
S_i \equiv e^{6\phi} J_i.
\eeqn
These variables together with $\rho_*$ and $\cB^i$
are evolved explicitly in the numerical simulation of the ideal MHD
(see Sec. II C). 

\subsection{Einstein's equation}

Our formulation for Einstein's equations is the same as 
in \cite{STU} in three spatial dimensions and in \cite{shiba2d} in
axial symmetry. Here, we briefly review the basic equations in our formulation. 
Einstein's equations are split into constraint and evolution equations. 
The Hamiltonian and momentum constraint equations are written as 
\beqn
&& R_k^{~k}- \tilde A_{ij} \tilde A^{ij}+
{2 \over 3} K^2=16\pi \rho_{\rm H},
\label{ham}\\
&& D_i \tilde A^i_{~j}-{2 \over 3}D_j K=8\pi J_j, \label{mom}
\eeqn
or, equivalently 
\beqn
&& \tilde \Delta \psi = {\psi \over 8}\tilde R_k^{~k} 
- 2\pi \rho_{\rm H} \psi^5 
-{\psi^5 \over 8} \Bigl(\tilde A_{ij} \tilde A^{ij}
-{2 \over 3}K^2\Bigr), \label{hameq} \\
&& \tilde D_i (\psi^6  \tilde A^i_{~j}) - {2 \over 3} \psi^6 
\tilde D_j K = 8\pi J_j \psi^6, \label{momeq}
\eeqn
where $\psi\equiv e^{\phi}$. 
These constraint equations are solved to set initial conditions.
A method in the case of GRMHD is presented in Sec. IV. 

In the following of this subsection, we assume that Einstein's equations
are solved in the Cartesian coordinates $(x, y, z$) for simplicity. 
Although we apply the implementation described here to axisymmetric
issues as well as nonaxisymmetric ones, this causes no problem 
since Einstein's equations in axial symmetry can be solved using the
so-called Cartoon method in which an axisymmetric boundary condition
is appropriately imposed in the Cartesian coordinates 
\cite{alcu,shiba2d0,shiba2d}: In the Cartoon method, 
the field equations are solved only in the $y=0$ plane, and
grid points of $y=\pm \Delta x$ 
($\Delta x$ denotes the grid spacing in the uniform grid)
are used for imposing the axisymmetric boundary conditions.

We solve Einstein's evolution equations in our latest  
BSSN formalism \cite{BSSN,STU}. In this formalism, a 
set of variables ($\tilde \gamma_{ij}, \phi, \tilde A_{ij}, K, F_i$) 
are evolved. Here, we adopt an auxiliary variable
$F_i \equiv \delta^{jl}\pa_l \tilde \gamma_{ij}$ 
that is the one originally proposed and 
different from the variable adopted in \cite{DLSS} in which
$\pa_i \tilde \gamma^{ij}$ is used. Evolution equations for
$\tilde \gamma_{ij}$, $\phi$, $\tilde A_{ij}$, and $K$ are 
\beqn
&&(\pa_t - \beta^l \pa_l) \tilde \gamma_{ij} 
=-2\alpha \tilde A_{ij} 
+\tilde \gamma_{ik} \beta^k_{~,j}+\tilde \gamma_{jk} \beta^k_{~,i}
-{2 \over 3}\tilde \gamma_{ij} \beta^k_{~,k}, \label{heq} \\
&&(\pa_t - \beta^l \pa_l) \tilde A_{ij} 
= e^{ -4\phi } \biggl[ \alpha \Bigl(R_{ij}
-{1 \over 3}e^{4\phi}\tilde \gamma_{ij} R_k^{~k} \Bigr) 
-\Bigl( D_i D_j \alpha - {1 \over 3}e^{4\phi}
\tilde \gamma_{ij} \Delta \alpha \Bigr)
\biggr] \nonumber \\
&& \hskip 2.5cm +\alpha (K \tilde A_{ij} 
- 2 \tilde A_{ik} \tilde A_j^{~k}) 
+\beta^k_{~,i} \tilde A_{kj}+\beta^k_{~,j} 
\tilde A_{ki}
-{2 \over 3} \beta^k_{~,k} \tilde A_{ij} \nonumber \\
&& \hskip 2.5cm-8\pi\alpha \Bigl( 
e^{-4\phi} S_{ij}-{1 \over 3} \tilde \gamma_{ij} S_k^{~k}
\Bigr), \label{aijeq} \\
&&(\pa_t - \beta^l \pa_l) \phi = {1 \over 6}\Bigl( 
-\alpha K + \beta^k_{~,k} \Bigr), \label{peq} \\
&&(\pa_t - \beta^l \pa_l) K 
=\alpha \Bigl[ \tilde A_{ij} \tilde A^{ij}+{1 \over 3}K^2
\Bigr] 
-\Delta \alpha +4\pi \alpha (\rho_{\rm H}+ S_k^{~k}). 
\label{keq}
\eeqn
For a solution of $\phi$, the following conservative form may be adopted
\cite{STU}: 
\beq
\pa_t e^{6\phi} -\pa_i (\beta^i e^{6\phi})= 
-\alpha K e^{6\phi}. \label{peq2} 
\eeq

For computation of $R_{ij}$ in the evolution equation of $\tilde A_{ij}$,
we decompose 
\beq
R_{ij}=\tilde R_{ij}+R^{\phi}_{ij},
\eeq
where 
\beqn
&&R^{\phi}_{ij}=-2\tilde D_i \tilde D_j \phi 
- 2  \tilde \gamma_{ij}\tilde \Delta \phi 
+ 4 \tilde D_i \phi \tilde D_j \phi 
- 4 \tilde \gamma_{ij} \tilde D_k \phi \tilde D^k \phi, \\ 
&&\tilde R_{ij}={1 \over 2}\biggl[
\delta^{kl}(-h_{ij,kl}+h_{ik,lj}+h_{jk,li})
+2\pa_k( f^{kl} \tilde \Gamma_{l,ij} ) 
-2 \tilde \Gamma^l_{kj}\tilde \Gamma^k_{il}\biggr].\label{eqij}
\eeqn
In Eq. (\ref{eqij}),  we split $\tilde \gamma_{ij}$ and $\tilde \gamma^{ij}$  
as $\delta_{ij}+h_{ij}$ and $\delta^{ij}+f^{ij}$, respectively. 
$\tilde \Gamma^k_{ij}$ is the Christoffel symbol with respect 
to $\tilde \gamma_{ij}$, and 
$\tilde \Gamma_{k,ij}=\tilde \gamma_{kl} \tilde \Gamma^l_{ij}$. 
Because of the definition det$(\tilde \gamma_{ij})=1$
(in the Cartesian coordinates), we use $\tilde \Gamma^k_{ki}=0$.

In addition to a flat Laplacian of $h_{ij}$, 
$\tilde R_{ij}$ involves terms linear in $h_{ij}$ as 
$\delta^{kl} h_{ik,lj} + \delta^{kl} h_{jk,li}$. 
To perform numerical simulation stably, we replace these terms
by $F_{i,j}+F_{j,i}$. This is the most important part in the
BSSN formalism, pointed out originally by Nakamura \cite{NOK}.
The evolution equation of
$F_i$ is derived by substituting Eq. (\ref{heq}) into the
momentum constraint as 
\beqn
(\pa_t - \beta^l \pa_l)F_i& =&-16\pi \alpha J_i+2\alpha 
\Bigl\{ f^{kj} \tilde A_{ik,j}
+f^{kj}_{~~,j} \tilde A_{ik} 
-{1 \over 2} \tilde A^{jl} h_{jl,i} 
+6\phi_{,k} \tilde A^k_{~i}-{2\over 3}K_{,i} \Bigr\} 
\nonumber \\
&+&\delta^{jk} \Bigl\{ -2\alpha_{,k} \tilde A_{ij} 
+ \beta^l_{~,k}h_{ij,l} 
+\Bigl(\tilde \gamma_{il}\beta^l_{~,j}+\tilde \gamma_{jl}\beta^l_{~,i}
-{2\over 3}\tilde \gamma_{ij} \beta^l_{~,l}\Bigr)_{,k}\Bigr\}. 
\label{eqF}
\eeqn

We also have two additional notes for handling  
the evolution equation of $\tilde A_{ij}$. One is on the method for 
evaluation of $R_k^{~k}$ for which there are two options,
use of the Hamiltonian constraint and direct calculation by 
\beqn
R_{ij}\gamma^{ij}=e^{-4\phi} (\tilde R_{k}^{~k} 
+R^{\phi}_{ij} \tilde \gamma^{ij}). 
\eeqn
We always adopt the latter one 
since with this, the conservation of the relation 
$\tilde A_{ij} \tilde \gamma^{ij}=0$ is much better preserved.
The other is on the handling of 
a term of $\tilde \gamma^{ij}\delta^{kl}h_{ij,kl}$
which appears in $\tilde R_k^{~k}$. This term is written by 
\beq
\tilde \gamma^{ij}\delta^{kl}h_{ij,kl}=
-\delta^{kl}h_{ij,k}f^{ij}_{~~,l}, \label{eq22}
\eeq
where we use ${\rm det}(\tilde \gamma_{ij})=1$
(in the Cartesian coordinates). 

As the time slicing condition, an approximate maximal slice 
condition $K \approx 0$ is adopted following previous papers (e.g.,
\cite{bina2}).  As the spatial gauge condition, we adopt a hyperbolic
gauge condition as in \cite{S03,STU}. Successful numerical results for
merger of binary neutron stars and stellar core collapse in these
gauge conditions are presented in \cite{STU,STU2,SS,SS3d}.
We note that these are also different from those in \cite{DLSS}. 

\subsection{GRMHD equations} \label{sec:grmhd}

Hydrodynamic equations are composed of
\beqn
&&\nabla_{\mu} (\rho u^{\mu})=0, \label{hyeq1}\\
&&\gamma_{i}^{~\nu} \nabla_{\mu} T^{\mu}_{~\nu}=0, \label{hyeq2}\\ 
&&n^{\nu} \nabla_{\mu} T^{\mu}_{~\nu}=0. \label{hyeq3}
\eeqn
The first, second, and third equations are the
continuity, Euler, and energy equations, respectively.
In the following, the equations are described for general
coordinate systems since the hydrodynamic equations are solved
in the cylindrical coordinates as well as in the Cartesian coordinates. 

The continuity equation (\ref{hyeq1}) is immediately written to 
\beqn
&& \pa_t \rho_*  + {1 \over \sqrt{\eta}}
\pa_i (\rho_*\sqrt{\eta} v^i)=0.\label{cont}
\eeqn
Equations (\ref{hyeq2}) and (\ref{hyeq3}) are rewritten as
\beqn
&&\pa_{\mu} (\sqrt{-g}T^{\mu}_{~i})
-{\sqrt{-g} \over 2}T^{\mu\nu}\pa_i g_{\mu\nu}=0,\\
&&\pa_{\mu} (\sqrt{-g}T^{\mu}_{~\nu}n^{\nu})
-\sqrt{-g} T^{\mu\nu}\nabla_{\mu}n_{\nu}=0. 
\eeqn
Then, using Eq. (\ref{31t}), they are written to 
\beqn
&& \pa_{\mu}[ \sqrt{-g} (n^{\mu} J_j + \gamma^{\mu i} S_{ij})]
=\sqrt{\gamma} \biggl(-\rho_{\rm H} \pa_j \alpha +J_i \pa_j \beta^i
-{\alpha \over 2} S_{ik}\pa_j \gamma^{ik}\biggr),\label{hyeq32} \\
&& \pa_{\mu}[ \sqrt{-g} (\rho_{\rm H} n^{\mu} + \gamma^{\mu i} J_{i})]
=\sqrt{\gamma} \biggl(\alpha K^{ij}S_{ij} -J_i D^i \alpha \biggr),
\label{hyeq33}
\eeqn
where we use
\beqn
&&n^{\mu} n^{\nu} \pa_j g_{\mu\nu} =-2 \pa_j \ln \alpha,\\
&&n^{\mu} \gamma^{\nu}_{~i} \pa_j g_{\mu\nu} =\alpha^{-1}\gamma_{ik}
\pa_j \beta^k,\\
&& \gamma^{\mu}_{~i} \gamma^{\nu}_{~k} \pa_j g_{\mu\nu}
=\pa_j \gamma_{ik},\\
&& \nabla_{\mu} n_{\nu}=-K_{\mu\nu}-n_{\mu} D_{\nu} \ln \alpha.
\eeqn
The explicit forms of Eqs. (\ref{hyeq32}) and (\ref{hyeq33}) are
\beqn
&& \pa_t S_j 
+ {1 \over \sqrt{\eta}}\pa_i \Bigl[\sqrt{\eta}
\Bigl\{ S_j v^i +\alpha e^{6\phi} \Ptot \delta_j^{~i}
-{\alpha \over w^2 e^{6\phi}} \cB^i(\cB_j+u_j \cB^k u_k)
\Bigr\} \Bigr] \nonumber \\
&&~~~~~~~=-S_0 \pa_j \alpha + S_k \pa_j \beta^k
+\alpha e^{6\phi}\Big[ 2 S_k^{~k} \pa_j \phi
+ \Ptot \pa_j \ln \sqrt{\eta}\Big]
-{1 \over 2} \alpha e^{2\phi}\hat S_{ik} \pa_j \tilde \gamma^{ik}, 
\label{euler2}\\
&& \pa_t S_0 + {1 \over \sqrt{\eta}}\pa_i \Bigl[\sqrt{\eta}\Big\{ S_0 v^i
+e^{6\phi} \Ptot (v^i+\beta^i)-{\alpha \over w e^{6\phi}}(\cB^k u_k) \cB^i
\Big\} \Bigr] \nonumber \\
&&~~~~~~~={1 \over 3} \alpha e^{6\phi} K S_k^{~k}+ \alpha e^{2\phi} \hat S_{ij} \tilde A^{ij} -S_k D^k \alpha,
\label{energy2}
\eeqn
where
\beqn
\hat S_{ij}=S_{ij}-\Ptot \gamma_{ij}. 
\eeqn

In the axisymmetric case, the equations for $(\rho_*, S_i, S_0)$
should be written in the cylindrical coordinates $(\varpi, \varphi, z)$
when we adopt the Cartoon method for solving Einstein's evolution
equations \cite{alcu,shiba2d0,shiba2d}. 
On the other hand, in the standard Cartoon method,
Einstein's equations are solved 
in the $y=0$ plane for which $x=\varpi$, $S_{\varpi}=S_x$, 
$S_{\varphi}=xS_y$, and other similar relations hold for vector and
tensor quantities. Taking into this fact, the hydrodynamic equations 
in axisymmetric spacetimes may be written using the Cartesian coordinates
replacing $(\varpi, \varphi)$ by $(x, y)$. Then,  
\beqn
&&\pa_t \rho_* + {1 \over x}\pa_x (\rho_* v^x x)+\pa_z (\rho_* v^z)=0,
\label{conti2d}\\
&&\pa_t S_A
+ {1 \over x}\pa_x \Big[ x \Big\{S_A v^x + \alpha e^{6\phi}
\Ptot \delta_A^{~x}
-{\alpha \over w^2 e^{6\phi}}\cB^x \Big(\cB_A+ u_A \cB^i u_i\Big)\Big\}\Big]
\nonumber \\
&&~~~~~~~~~~~+ \pa_z \Big[ S_A v^z + \alpha e^{6\phi} \Ptot \delta_A^{~z}
-{\alpha \over w^2 e^{6\phi}}\cB^z (\cB_A+ u_A \cB^i u_i)\Big]
\nonumber \\
&&~~~~~~~~~~~=-S_0 \pa_A \alpha + S_k \pa_A \beta^k 
+\alpha e^{6\phi}\Big[ 2 S_k^{~k} \pa_A \phi 
+ {\Ptot \over x}\delta_{A}^{~x} \Big]
-{1 \over 2} \alpha e^{2\phi}\hat S_{ik} \pa_A \tilde \gamma^{ik}
\nonumber \\
&&~~~~~~~~~~~~+\Big[
{S_y v^y \over x}-{\alpha \over x w^2 e^{6\phi}}\cB^y (\cB_y +\cB^i u_i u_y)
\Big]\delta_A^{~x},
\\
&&\pa_t S_y + {1 \over x^2}\pa_x \Big[x^2
\Big\{S_y v^x-{\alpha \over w^2 e^{6\phi}}
\cB^x\Big(\cB_y + u_y \cB^i u_i \Big)\Big\}\Big] \nonumber \\
&&~~~~~~~~~~~~
+\pa_z \Big[S_y v^z-{\alpha \over w^2 e^{6\phi}}
\cB^z \Big(\cB_y + u_y \cB^i u_i \Big) \Big]=0,\\
&&\pa_t S_0
+ {1 \over x}\pa_x \Big[x\Big\{ S_0 v^x + e^{6\phi}\Ptot (v^x + \beta^x)
-{\alpha \over w e^{6\phi}}\cB^i u_i \cB^x \Big\} \Big]
+ \pa_z \Big[ S_0 v^z + e^{6\phi}\Ptot (v^z + \beta^z)
-{\alpha \over w e^{6\phi}}\cB^i u_i \cB^z \Big] \nonumber \\
&&~~~~~~~~~~~~={1 \over 3} \alpha e^{6\phi} K S_k^{~k}+ \alpha e^{2\phi}
\hat S_{ij} \tilde A^{ij} -S_k D^k \alpha, \label{ene2d}
\eeqn
where $A$ denotes $x$ or $z$, while
$i, j, k, \cdots$ are $x$ or $y$ or $z$.

After evolving $\rho_*$, $S_i$, and $S_0$ together
with $\cB^i$ (see next subsection for the equations), we have to determine
the primitive variables such as $\rho$, $\varep$, $u_i$, and $u^t$
(or $w=\alpha u^t$).
For this procedure, we make an equation from the definition of
$S_i$ as 
\beqn
s^2 \equiv \rho_*^{-2} \gamma^{ij} S_i S_j =\Big(h + B^2 w^{-1}\Big)^2(w^2-1)
-D^2 (h w)^{-2}(B^2 + 2 h w), \label{newton1}
\eeqn
where $B^2$ and $D^2$ are determined from the evolved variables
$(\rho_*, S_i, \cB^i, \phi)$ as 
\beqn
B^2 = {\cB^2 \over \rho_* e^{6\phi}}~~~{\rm and}~~~
D^2 = {(\cB^i S_i )^2 \over \rho_*^3 e^{6\phi}},
\eeqn
and for getting Eq. (\ref{newton1}), 
we use the relation $S_i \cB^i =\rho_* h \cB^i u_i$. 
Equation (\ref{newton1}) is regarded as a function of $h$ and $w$
for given data sets of $s^2$, $B^2$, and $D^2$.
From the definition of $S_0$, we also make a function of $h$ and $w$ as 
\beqn
{S_0 \over \rho_*}=h w - {P \over \rho w}+B^2
-{1 \over 2w^2}(B^2 + D^2 h^{-2}). \label{newton2}
\eeqn
Here, $P/\rho$ may be regarded as a function of $h$ and $w$
for a given data sets of $\rho_*$ and $S_0$. This is indeed the case 
for frequently used equations of state such as 
$\Gamma$-law equations of state $P=(\Gamma-1)\rho\varep$
where $\Gamma$ is an adiabatic constant and hybrid equations of state
for which $P$ is written in the form $P(\rho, h)$ (e.g., \cite{HD,STU2}). 
Thus, Eqs. (\ref{newton1}) and (\ref{newton2}) constitute
simultaneous equations for $h$ and $w$ for given values of 
$\rho_*$, $S_i$, $S_0$, $\cB^i$, and geometric variables.
The solutions for $h$ and $w$
are numerically computed by the Newton-Raphson method very simply.
Typically, a convergent solution is obtained with four iterations
according to our numerical experiments. 

\subsection{Maxwell equations}

The Maxwell equations are 
\beqn
&&\nabla_{\mu} F^{\mu\nu}=-4\pi j^{\nu},\label{em1} \\
&&\nabla_{\mu} F_{\alpha\beta}+\nabla_{\alpha} F_{\beta\mu}
+\nabla_{\beta} F_{\mu\alpha}=0. \label{em2}
\eeqn
In the ideal MHD, Eq. (\ref{em1}) is not necessary, and
only Eq. (\ref{em2}) has to be solved. 
Using the dual tensor, Eq. (\ref{em2}) is rewritten to
\beqn
\nabla_{\mu} F^{*~\mu}_{~\nu}=0. 
\eeqn
This immediately leads to
\beqn
&& \pa_k(\eta^{1/2} \cB^k) = 0, \label{emcon1} \\
&& \pa_t  \cB^k = {1 \over \eta^{1/2}}\pa_i \Bigl[\eta^{1/2}
(\cB^i v^k - \cB^k v^i)\Bigr]. 
\label{emcon2}
\eeqn
Equation (\ref{emcon1}) is the no-monopoles constraint, and
Eq. (\ref{emcon2}) is the induction equation.
The constraint equation (\ref{emcon1}) is solved
in giving initial conditions, and the induction equation is 
solved for the evolution. 

In the axisymmetric case, these equations in the $y=0$ plane are
written as
\beqn
&& {1 \over x} \pa_x(x \cB^x) + \pa_z \cB^z= 0, \label{mag2d1}\\
&& \pa_t  \cB^x =  - \pa_z (\cB^x v^z - \cB^z v^x),\label{mag2d2}\\ 
&& \pa_t  \cB^z = {1 \over x}\pa_x \Big[x
(\cB^x v^z - \cB^z v^x)\Big],\\ 
&& \pa_t  \cB^y =  \pa_x (\cB^x v^y - \cB^y v^x)
+\pa_z (\cB^z v^y - \cB^y v^z). \label{mag2d4}
\eeqn
Equations (\ref{mag2d2})--(\ref{mag2d4}) together with 
Eqs. (\ref{conti2d})--(\ref{ene2d}) constitute basic equations for 
ideal MHD in the axisymmetric case. 

\subsection{Definition of global quantities}

In numerical simulations for self-gravitating system,
in addition to the total baryon rest mass $M_*$, we refer to 
the ADM mass and the angular momentum of the system, which are given by 
\beqn
M &&\equiv -{1 \over 2\pi} 
\oint_{r\rightarrow\infty} \pa_i \psi dS_i \nonumber \\
&&=\int \biggl[ \rho_{\rm H} e^{5\phi} +{e^{5\phi} \over 16\pi}
\biggl(\tilde A_{ij} \tilde A^{ij}-{2 \over 3}K^2 
-\tilde R_k^{~k} e^{-4\phi}\biggr)\biggr]d^3x, \label{eqm00}\\
J &&\equiv {1 \over 8\pi}\oint_{r\rightarrow\infty} 
\varphi^ i \tilde A_i^{~j} e^{6\phi} dS_j \nonumber \\
&&=\int e^{6\phi}\biggl[S_i \varphi^i  
+{1 \over 8\pi}\biggl( \tilde A_i^{~j} \pa_j \varphi^i
-{1 \over 2}\tilde A_{ij}\varphi^k\pa_k \tilde \gamma^{ij}
+{2 \over 3}\varphi^j \pa_j K \biggr) \biggr]d^3x,
\label{eqj00}
\eeqn
where $dS_j=r^2 \pa_j r d(\cos\theta)d\varphi$, 
$\varphi^j=-y(\pa_x)^j + x(\pa_y)^j$, and $\psi=e^{\phi}$.
In this paper, simulations are performed
in axial symmetry, and hence, $J$ is conserved. $M$ is
approximately conserved since the emission of 
gravitational waves is negligible. Thus, 
conservation of these quantities is checked during numerical
simulations.

The violation of the Hamiltonian constraint 
is locally measured by the equation as 
\beqn
\displaystyle
f_{\psi} &&\equiv 
\Bigl|\tilde \Delta \psi - {\psi \over 8}\tilde R_k^{~k} 
+ 2\pi \rho_{\rm H} \psi^5
+{\psi^5 \over 8} \Bigl(\tilde A_{ij} \tilde A^{ij}
-{2 \over 3}K^2\Bigr)\Bigr|  \nonumber \\
&&\times \biggl[|\tilde \Delta \psi | + |{\psi \over 8}\tilde R_k^{~k}| 
+ |2\pi \rho_{\rm H} \psi^5|
+{\psi^5 \over 8} \Bigl(\tilde A_{ij} \tilde A^{ij}+
{2 \over 3}K^2\Bigr)\biggr]^{-1}. 
\eeqn
Following \cite{shiba2d}, we define and monitor a global quantity as 
\beq
H \equiv {1 \over M_*} \int \rho_* f_{\psi} d^3x. \label{vioham}
\eeq
Hereafter, this quantity will be referred to as the averaged violation
of the Hamiltonian constraint.

\section{Numerical scheme for solving GRMHD equations} 

\subsection{GRMHD equations}

As described in Sec. \ref{sec:grmhd},
we write the GRMHD equations in the conservative form.
In this case, roughly speaking, there are two options for 
numerically handling the transport terms \cite{MM}.
One is to use the Godunov-type, approximate Riemann
solver \cite{komissarov,balsara}, and the other is to use the
high-resolution central (HRC) scheme \cite{delzanna,gammie}.
We adopt a HRC scheme proposed by Kurganov and Tadmor \cite{KT} 
and very recently used in special relativistic simulations 
by Lucas-Sarrano et al. \cite{lucas}. Thus our numerical
scheme for a solution of GRMHD equations is slightly different from that
in \cite{DLSS}, in which the HLL scheme \cite{HLL} is basically adopted. 

The basic equations can be schematically written as
\beqn
{\pa {\bf U} \over \pa t}+{\pa {\bf F}^i \over \pa x^i}+
{\pa \ln \sqrt{\eta} \over \pa x^i} {\bf F}^i={\bf S},
\eeqn
where
\beqn
&&{\bf U}=(\rho_*, S_i, S_0, \cB^i),\\
&&{\bf F}^j=(\rho_* v^j, 
S_i v^j+ \alpha e^{6\phi} \Ptot \delta_i^{~j}-\tau_{~i}^{B~j}, 
S_0 v^j+ e^{6\phi}\Ptot (v^j+\beta^j) - \tau_{~0}^{B~j},
\cB^i v^j-\cB^j v^i), 
\eeqn
and ${\bf S}$ denotes the terms associated with the gravitational force. 
Here, $\tau_{\mu}^{B~j}$ denotes a magnetic stress defined by 
\beqn
&& \tau_{~i}^{B~j}
={\alpha \over w^2 e^{6\phi}}\cB^j [\cB_i + u_i (\cB^k u_k)],\\
&& \tau_{~0}^{B~j}
={\alpha \over w e^{6\phi}}(\cB^k u_k)\cB^j. 
\eeqn
In addition to ${\bf U}$, we define a set of variables as 
\beqn
{\bf P}=(\rho_*, \hat u_i, \varepsilon, \cB^i). 
\eeqn
$\hat u_i$ and $\varepsilon$ are computed at each time step
from Eqs. (\ref{newton1}) and (\ref{newton2}). We use ${\bf P}$
for the reconstruction of ${\bf F}$ at cell interfaces.
In standard method, one often uses a set of primitive variables
$(\rho, v^i, \varep, \cB^i)$ instead of ${\bf P}$ for reconstruction
of ${\bf F}$. We have found in the test problems
that even using $\rho_*$ and $\hat u_i$
instead of $\rho$ and $v^i$, it is possible to guarantee the similar
accuracy and stability. 

To evaluate ${\bf F}$, we use a HRC scheme \cite{lucas}.
The fluxes are defined at cell faces. A piece-wise parabolic
interpolation from the cell centers gives ${\bf P}_R$ and
${\bf P}_L$, the primitive variables at the right- and left-hand
side of each cell interface, as
\beqn
&&Q_L=Q_i
+{\Phi(r^+_{i-1})\Delta_{i-1} \over 6}+{\Phi(r^-_{i})\Delta_{i} \over 3},\\
&&Q_R=Q_{i+1}
-{\Phi(r^+_{i})\Delta_i \over 3}-{\Phi(r^-_{i+1})\Delta_{i+1} \over 6}.
\eeqn
Here, $Q$ denotes a component of ${\bf P}$ 
and $\Delta_{i+1}=Q_{i+1}-Q_i$. 
$\Phi$ denotes a limiter function defined by 
\beqn
\Phi(r)={\rm minmod}(1, br) ~~~~~(1\leq b \leq 4~{\rm for~TVD~condition}),
\eeqn
where $r^{\pm}_i=\Delta_{i \pm 1}/\Delta_i$, and
\beqn
{\rm minmod}(1, x)=\left\{
\begin{array}{ll}
1 & {\rm if}~~x>1 \\
x & {\rm if}~~1>x>0 \\
0 & {\rm if}~~x<0
\end{array}
\right.
.
\eeqn
For the simulations presented in Secs. V and VI,
we choose $b=2$ unless otherwise
stated. We have found that the dissipation is relatively large for
$b=1$ with which it is difficult to evolve isolated neutron stars
for a long time scale accurately. On the other hand,
for $b \geq 3$, the dissipation is
so small that instabilities often occur around strong discontinuities,
and around the region for which $\Ptot \gg P$. 

From ${\bf P}_L$ and ${\bf P}_R$, we calculate the maximum wave
speed $c_L$ and $c_R$, and the fluxes ${\bf F}_L$ and ${\bf F}_R$
at the right- and left-hand sides of each cell interface. Then,
we define $c_{\rm max}={\rm max}(c_L, c_R)$, and the flux
\beqn
{\bf F}={1 \over 2}\biggl[{\bf F}_L+{\bf F}_R
-c_{\rm max}({\bf U}_R-{\bf U}_L)\biggr]. 
\eeqn

In adopting the central schemes, the eigen vectors for the Jacobi
matrix $\pa {\bf F}/\pa {\bf U}$ are not required in contrast to the
case with the Godunov-type scheme \cite{MM,ref9}. However, the eigen
values for each direction are still necessary to evaluate
characteristic wave speeds $c_L$ and $c_R$.
The equation for the seven eigen values $\lambda$ is derived by
Anile and Pennisi \cite{anile}: Three of the seven solutions for
$\lambda$ in $x^i$ direction are described by 
\beqn
\lambda=v^i,~~~
{b^i \pm u^i \sqrt{\rho h + b^2} \over b^t \pm u^t \sqrt{\rho h + b^2}}, 
\eeqn
and rest four are given by the solutions
for the following fourth order equation 
\beqn
&&(u^t)^4(\lambda-v^i)^4(1-\zeta)+\Bigl[
c_s^2{(b^i-\lambda b^t)^2 \over \rho h + b^2}-(u^t)^2(\lambda-v^i)^2
\Big(\gamma^{ii}-{\beta^i+\lambda \over \alpha^2}\Big)\zeta\Bigr]=0
~~({\rm no~summation~for}~i). \label{fourth}
\eeqn
Here, $\zeta$, the sound velocity $c_s$, and the Alfv\'en velocity $v_A$
are defined, respectively, by 
\beqn
&&\zeta \equiv v_A^2+ c_s^2 -v_A^2 c_s^2, \\
&&c_s^2 \equiv {1 \over h}\biggl[{\pa P \over \pa \rho}\Big|_{\varep}
+{P \over \rho^2} {\pa P \over \pa \varep}\Big|_{\rho}\biggr],\\
&& v_A^2 \equiv {b^2 \over \rho h + b^2}. 
\eeqn
In the central schemes, we only need the maximum characteristic speed,
and thus, only the solutions for Eq. (\ref{fourth}), which contain the
fast mode, are relevant. The solutions for the fourth-order equation 
are determined either analytically or by standard numerical methods.
However, for simplicity and for saving computational time, 
we use the prescription proposed by Gammie et al. \cite{gammie}, who
have found it convenient to replace the fourth-order equation 
approximately by a second-order one:
\beqn
(u^t)^2(\lambda-v^i)^2(1-\zeta) - \zeta \Big( 
\gamma^{ii}-{\beta^i+\lambda \over \alpha^2}\Big)=0~~
({\rm no~summation~for}~i). \label{second}
\eeqn
The solution of Eq. (\ref{second})
for an arbitrary direction $x^i$ is written as
\beqn
\lambda^i={1 \over \alpha^2 - V_k V^k \zeta}
&& \biggl[v^i \alpha^2 (1-\zeta) -\beta^i \zeta (\alpha^2 - V^2)
\nonumber \\
&&~~ \pm \alpha \sqrt{\zeta} \sqrt{(\alpha^2 -V^2)\{ \gamma^{ii}
(\alpha^2 - V^2 \zeta)-(1-\zeta)V^i V^i \}} \biggr]
~~({\rm no ~summation ~for}~ i), 
\eeqn
where $V^i=v^i+\beta^i$ and $V^2=\gamma_{ij}V^i V^j$. 
This is equivalent to that obtained by replacing $c_s$ by $\sqrt{\zeta}$ 
in the solution for the pure hydrodynamic case \cite{Val,Toni,shiba2d}. 

\subsection{Induction equation}

The induction equation may be solved using the same scheme as
in solving the hydrodynamic equations described above.
However, with such a scheme, the violation of the
constraint equation (\ref{emcon1}) is often accumulated with time,
resulting in a nonreliable solution. Thus, we adopt a
constraint transport scheme \cite{EH}. Namely, we put the
components of the magnetic field at the cell-face centers.
Here, we specifically consider the axisymmetric case with the
cylindrical coordinates $(x, \varphi, z)$~($\varpi$ is
replaced by $x$). Extension to the
nonaxisymmetric case is straightforward, and the description below
can be used with slight modification.

In the axisymmetric case with the cylindrical coordinates,
the numerical computation is performed in a discretized cell for
$(x, z)$. Here, we denote the cell center for $(x, z)$ by $(i, j)$. 
Then, we put $\cB^x$ at $(i+1/2, j)$,
and $\cB^z$ at $(i, j+1/2)$ while components of the gravitational
field and fluid variables as well as $\cB^y \equiv x \cB^{\varphi}$
are put at the cell center $(i, j)$. In this case, the induction
equations for $\cB^x$ and $\cB^z$ are solved in a constraint transport
scheme \cite{EH}, while that for $\cB^y$ is solved in the same method as
that for the continuity equation of $\rho_*$. 

Computing the flux at cell edges for the induction equation is 
different from that for the fluid equation.
This is because numerical fluxes have to be defined so that the
constraint equation (\ref{emcon1}) is satisfied.  
For example, for the $x$ component of the induction equation, the flux
in the $z$ direction is written as $v^z \cB^x - v^x \cB^z \equiv F_1$.
On the other hand,
for the $z$ component of the induction equation, the flux in the
$x$ direction is written as $v^x \cB^z - v^z \cB^x \equiv F_2$. Both 
$F_1$ and $F_2$ have to be defined at cell edges $(i+1/2, j+1/2)$, and 
for the constraint equation (\ref{emcon1}) to be satisfied, we
have to require $F_1=-F_2=F$ at each cell edge. 
In addition, an upwind scheme should be adopted for numerical stability: 
For the induction equation of $\cB^x$, the upwind prescription should be
applied for the $z$ component of the flux. On the other hand,  
for the induction equation of $\cB^z$, the upwind prescription should
be applied for the $x$ component of the flux. $F$ has to be determined 
taking into account these requirements.

We here adopt a scheme proposed by Del Zanna et al. \cite{delzanna},
which satisfies such requirements. In this scheme, the flux is written as
\beqn
F = {1 \over 4}(F^{LL}+F^{LR}+F^{RL}+F^{RR})
-{c_{\rm max}^z \over 2} [(\cB^{x})^R-(\cB^{x})^L]
+{c_{\rm max}^x \over 2} [(\cB^{z})^R-(\cB^{z})^L], 
\eeqn
where, e.g., $F^{LR}$ is the flux defined at the left-hand side in the
$x$ direction and at right-hand side in the $z$ direction.
These fluxes are computed by a piece-wise parabolic interpolation. 
$c_{\rm max}^i$ is the characteristic speed for the prescription
of an upwind flux-construction 
and calculated at cell edges using the interpolated variables.
For simplicity, we set 
$c_{\rm max}^z={\rm max}(v^z_L, v^z_R)$ and
$c_{\rm max}^x={\rm max}(v^x_L, v^x_R)$.

For solving other equations, 
it is necessary to define the magnetic field at the cell center.
Since the $x$ and $z$ components of the magnetic field are defined at
the cell face centers (i.e., $\cB^x$ at $(i+1/2,j)$ and
$\cB^z$ at $(i, j+1/2)$), this is done by a simple averaging as
\beqn
&&\cB^x_{i,j}={1 \over 2}(\cB^x_{i+1/2,j}+\cB^x_{i-1/2,j}), \\
&&\cB^z_{i,j}={1 \over 2}(\cB^z_{i,j+1/2}+\cB^z_{i,j-1/2}). 
\eeqn
Also, $v^i$ at the cell face center
is necessary for computing $c_{\rm max}^x$ and $c_{\rm max}^z$.
To compute them, we also use a simple averaging. 
For the definition of $v^k_{i+1/2,j}$ and $v^k_{i,j+1/2}$, 
we have also tried the Roe-type averaging in terms of $\rho_*^{1/2}$, but 
any significant modification in the results has not been found. 

Before closing this section, we note that our scheme for the induction
equation is different from that adopted in \cite{DLSS}, in which
a T\'oth's method is used \cite{Toth}.

\section{Initial value problem} 

In the fully general relativistic and dynamical simulations,
we have to solve the constraint equations of general relativity
for preparing the initial condition. One solid method is 
to give an equilibrium state. For rigidly rotating stars of poloidal
magnetic fields in axial symmetry, such equilibrium has been
already computed \cite{BBGJ}. However, for differentially rotating
stars or nonaxisymmetric cases, the method has not been established.
Thus, we here present a simple method for preparing an 
initial condition which is similar to that in \cite{MO}. 
In the following, we assume that axisymmetric 
matter fields $\rho_*$, $\hat e\equiv h w-P/\rho w$, $h$, and
$\hat u_i\equiv h u_i$ are a priori given (e.g., those for
rotating stars of no magnetic field in equilibrium are given). 
Although we assume the axial symmetry, 
the same method can be applied for the nonaxisymmetric case. 

Initial conditions for magnetic fields have to satisfy
Eq. (\ref{emcon1}). A solution of Eq. (\ref{emcon1}) is written as
\beqn
\cB^k =e^{kij}\pa_i A_j, \label{vectorA}
\eeqn
where $A_j$ is an arbitrary vector potential and $e^{kij}$ is
a Levi-Civita tensor of flat three-space. If we choose
\beqn
A_{x}=A_{z}=0,~~{\rm and}~~A_{\varphi}\not=0,
\eeqn
the magnetic fields are poloidal. Here, we assume to use
the cylindrical coordinates $(x, \varphi, z)$~($\varpi$ is replaced by $x$). 

In the axisymmetric case, we can also choose pure toroidal magnetic fields
as 
\beqn
\cB^{x}=\cB^{z}=0,~~{\rm and}~~\cB^{\varphi}\not=0,
\eeqn
where $\cB^{\varphi}$ may be an arbitrary function.
In the following, we give a nonzero function
either for $A_{\varphi}$ or for $\cB^{\varphi}$. 

Initial conditions also have to satisfy Eqs. (\ref{hameq}) and (\ref{momeq}).
In the following, we assume that 
$\tilde \gamma_{ij}$ and $K$ are given functions in these equations. 
Remind that $\rho_{\rm H}$ and $J_i$ are written as
\beqn
&& \rho_{\rm H}=\rho_* \hat e \psi^{-6}+\psi^{-12}
\Big(\cB^2 - {\cB^2 + (\cB^i u_i)^2 \over 2 w^2}\Big),\\
&& J_i = \rho_* \hat u_i \psi^{-6}+ {1 \over w \psi^{12}}
\Big(\cB^2 u_i- \cB_i \cB^j u_j \Big),
\eeqn
where $\psi$ denotes the conformal factor $(=e^{\phi})$, and
$w=\sqrt{1+\psi^{-4} \tilde \gamma^{ij} u_i u_j}$. 
Thus, if $\rho_*$, $\hat e$, $h$, $\hat u_i$, 
$A_{\varphi}$, $\cB^{\varphi}$, $K$, and $\tilde \gamma_{ij}$ are given,
the remaining unknown functions are $\psi$
and $\tilde A_{ij}$. This implies that
the constraint equations are solved for these variables using
the technique developed by York \cite{york}.

First, we decompose the tracefree part of the extrinsic curvature as
\beqn
\hat A_{ij}\equiv \psi^6 \tilde A_{ij}
=\tilde D_i W_j + \tilde D_j W_i - {2 \over 3}\tilde \gamma_{ij}
\tilde D_k W^k + K_{ij}^{\rm TT}, \label{decompose}
\eeqn
where $W_i$ is a three vector, and $K_{ij}^{\rm TT}$ is 
a transverse-tracefree tensor which satisfies
\beqn
\tilde D^i K_{ij}^{\rm TT}=0=K_{ij}^{\rm TT} \tilde \gamma^{ij}. 
\eeqn
$K_{ij}^{\rm TT}$ would be composed mainly of gravitational waves. 
Hereafter, we set $K_{ij}^{\rm TT}=0$ for simplicity. 
Using Eq. (\ref{decompose}), Eq. (\ref{momeq}) is rewritten to
\beqn
&& \tilde \Delta W_j +{1 \over 3} \tilde D_j \tilde D_i W^i
+\tilde R_{ji} W^i - {2 \over 3} \psi^6 
\tilde D_j K = 8\pi J_i \psi^6. 
\label{momeq3} 
\eeqn
This equation can be solved for an initial trial function of $\psi$.
Then, $\hat A_{ij}$ is computed from Eq. (\ref{decompose}).
Substituting $\hat A_{ij}$, the Hamiltonian constraint (\ref{hameq})
is solved in the next step. Then we solve the momentum constraint
again, and repeat these procedures until a sufficient convergence
is achieved. 

\section{Special relativistic tests}

In this section, we present numerical results for
a number of special relativistic tests. In the tests, we
adopt the $\Gamma$-law equations of state as
\beqn
P=(\Gamma-1)\rho \varepsilon,
\eeqn
with $\Gamma=4/3$ or $5/3$.
Simulations are always performed using the uniform grid 
in all the axis directions. 

\subsection{One dimensional tests}

\begin{figure}[p]
\vspace{-4mm}
\begin{center}
\epsfxsize=3.in
\leavevmode
(a)\epsffile{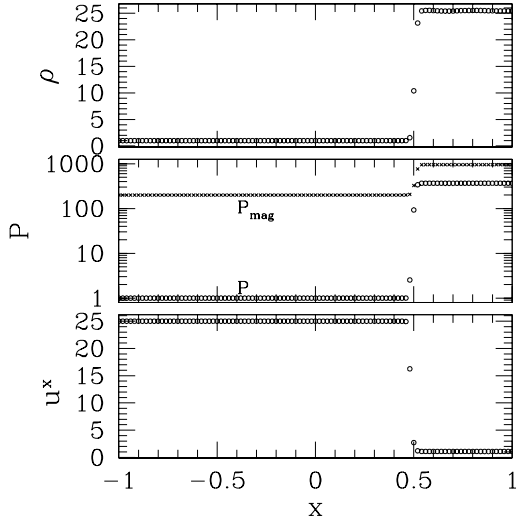} 
\epsfxsize=3.in
\leavevmode
~~~~(b)\epsffile{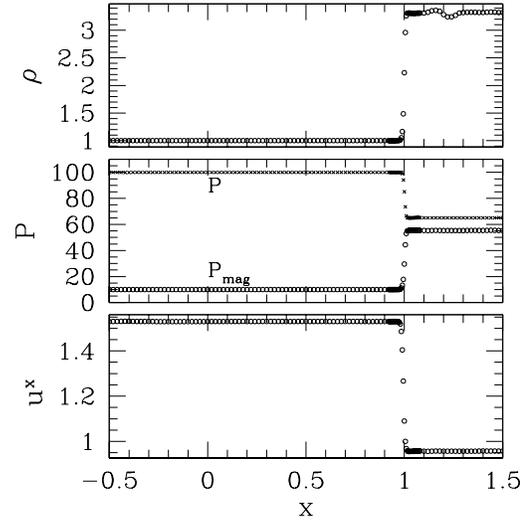} 
\end{center}
\vspace{-2mm}
\caption{(a) Propagation of a fast shock.
The snapshot at $t=2.5$ is shown. Numerical simulation is
performed with $N=100$ and $\Delta x=0.02$. 
(b) Propagation of a slow shock. The snapshot at $t=2$ is shown. The 
numerical simulation is performed with $N=400$ and $\Delta x=0.005$.
Only one fourth of data points are plotted except for [0.92,1.08] in which 
all the data points are plotted. 
For both cases, the initial discontinuities were located at $x=0$, and
the shock fronts move with a constant velocity $\mu$ where
$\mu=0.2$ and 0.5 for the fast and slow shocks, respectively. 
\label{FIG1} }
\end{figure}

\begin{figure}[p]
\vspace{-4mm}
\begin{center}
\epsfxsize=3.in
\leavevmode
(a)\epsffile{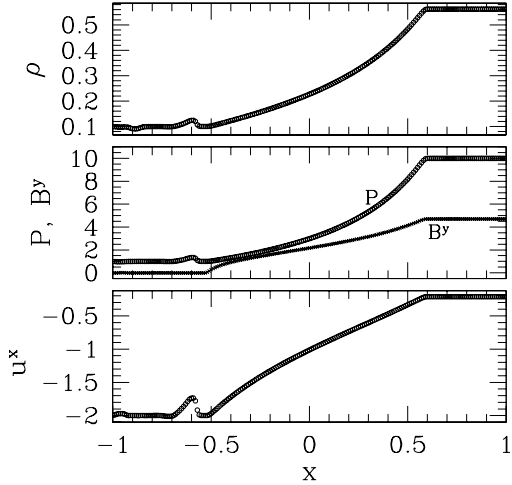} 
\epsfxsize=3.in
\leavevmode
~~~~(b)\epsffile{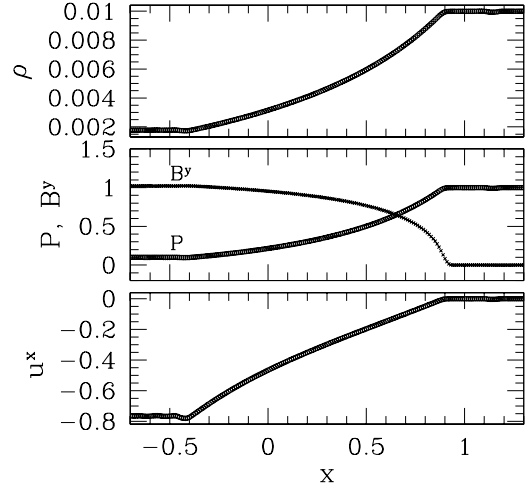} 
\end{center}
\vspace{-2mm}
\caption{(a) Propagation of a fast switch-off rarefaction waves.
The snapshot at $t=1.0$ is shown. 
(b) Propagation of a slow switch-on rarefaction waves.
The snapshot at $t=2$ is shown.
For both cases, the initial discontinuities were located at $x=0$, and
the numerical simulations are performed with $N=400$ and $\Delta x=0.005$.
Only half of all the data are plotted for both figures.
\label{FIG2} }
\end{figure}

\begin{figure}[p]
\vspace{-4mm}
\begin{center}
\epsfxsize=3.in
\leavevmode
(a)\epsffile{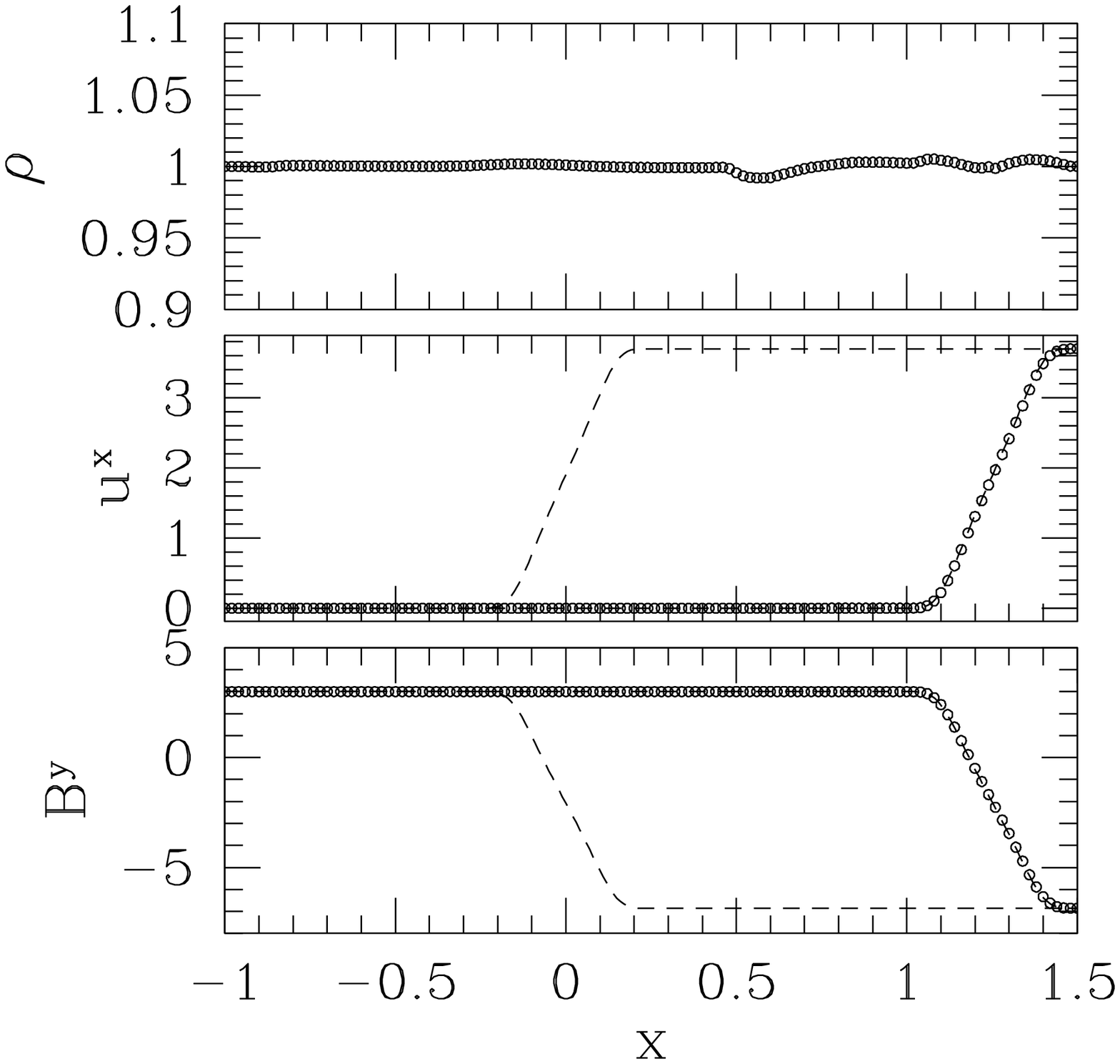} 
\epsfxsize=3.in
\leavevmode
~~~(b)\epsffile{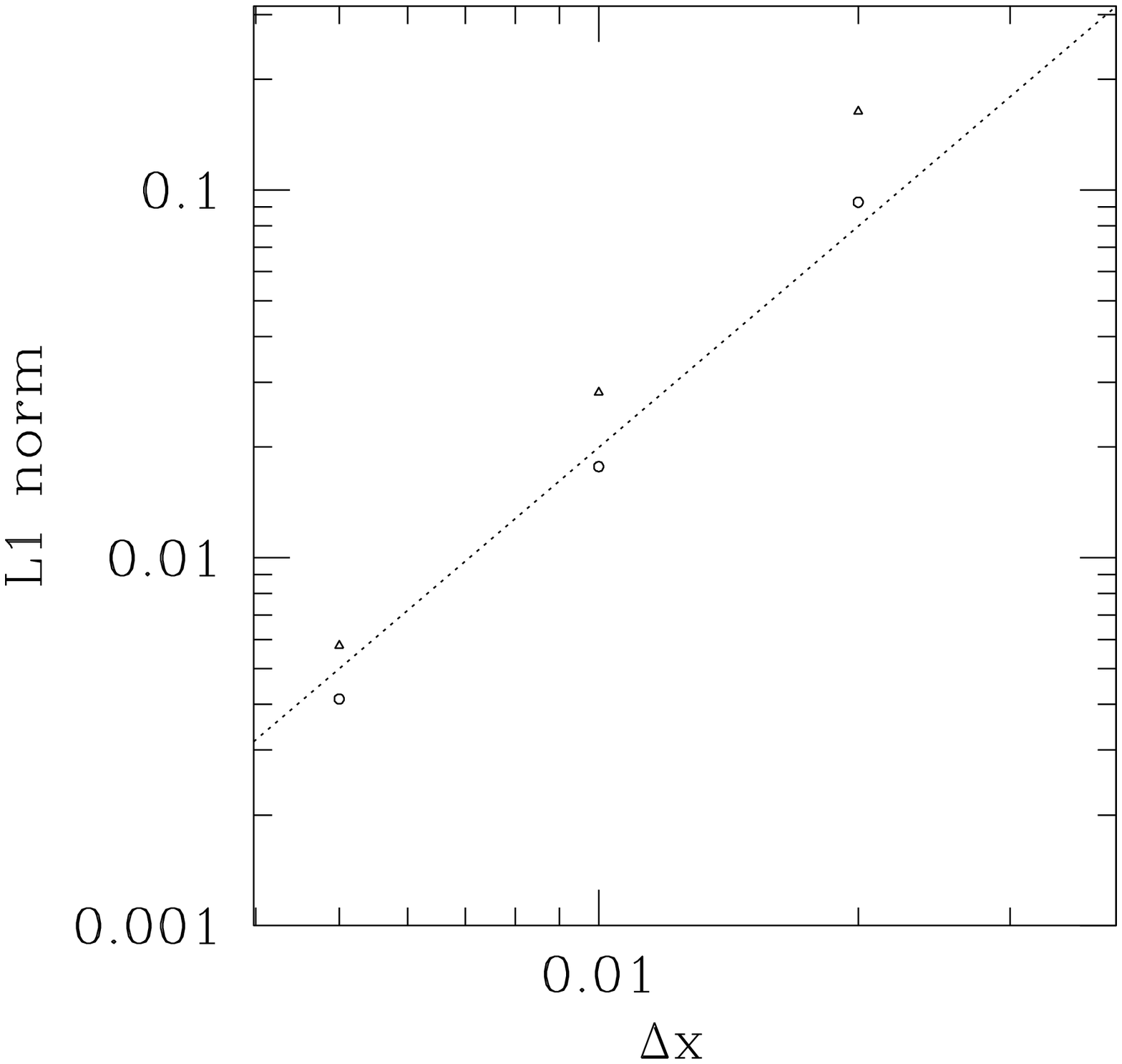} 
\end{center}
\vspace{-2mm}
\caption{(a) Propagation of a strong continuous Alfv\'en waves.
The snapshot at $t=2$ is shown. The initial configurations
are the analytic solution at $t=2$ are shown by the dashed curves. 
The numerical simulation is performed with $N=500$ and $\Delta x=0.005$.
Only 1/4 of all the data are plotted. 
(b) L1 norm of the error for $\rho$ (circles) and $P$ (triangles)
as a function of $\Delta x$. These variables should be constant
in this problem, and the deviation from the stationary values is
due to a numerical error. The dotted line denotes the
slope expected for second-order convergence. 
\label{FIG3} }
\end{figure}

\begin{figure}[p]
\begin{center}
\epsfxsize=3.in
\leavevmode
(a)\epsffile{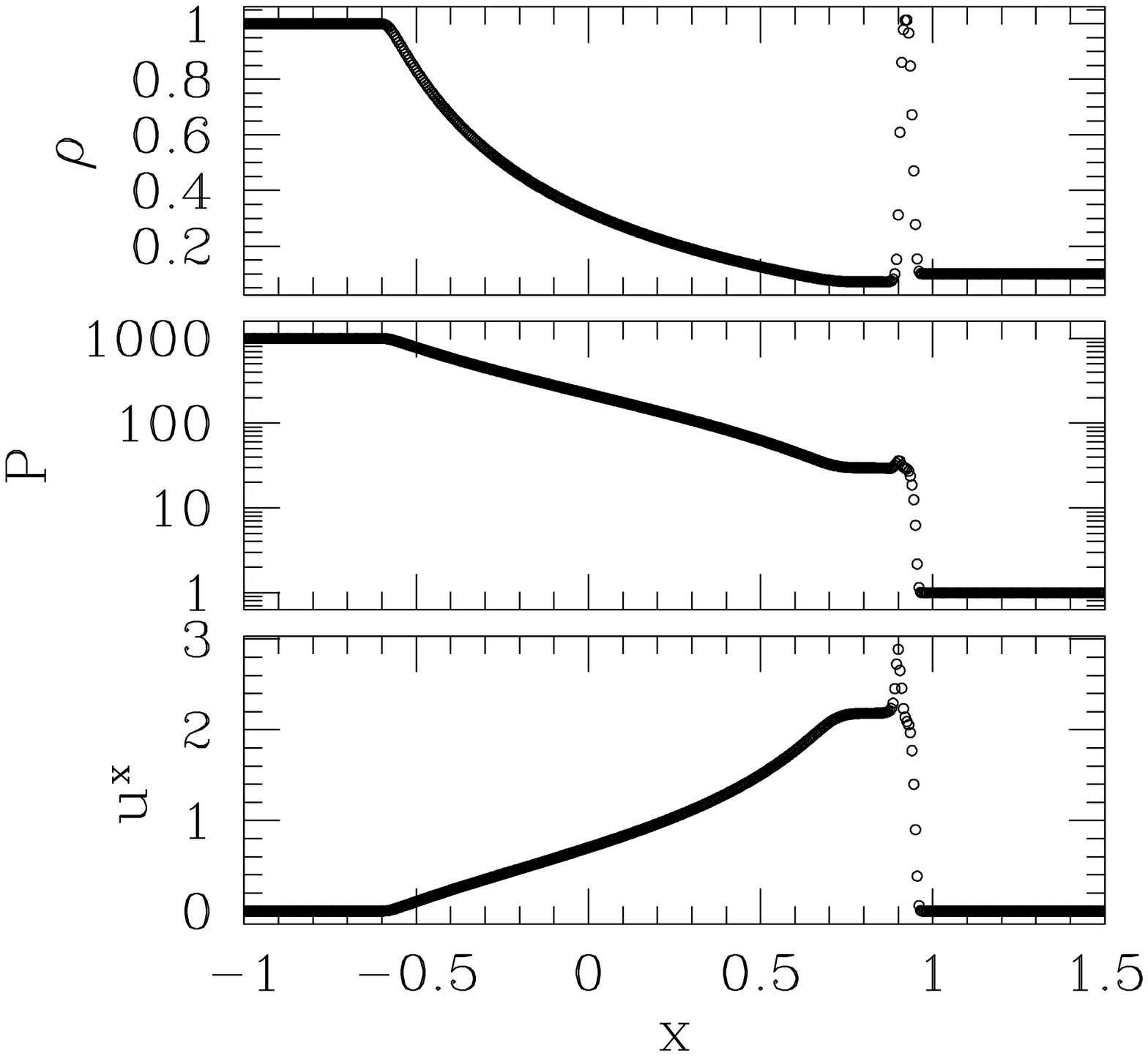} 
\epsfxsize=3.in
\leavevmode
~~~~(b)\epsffile{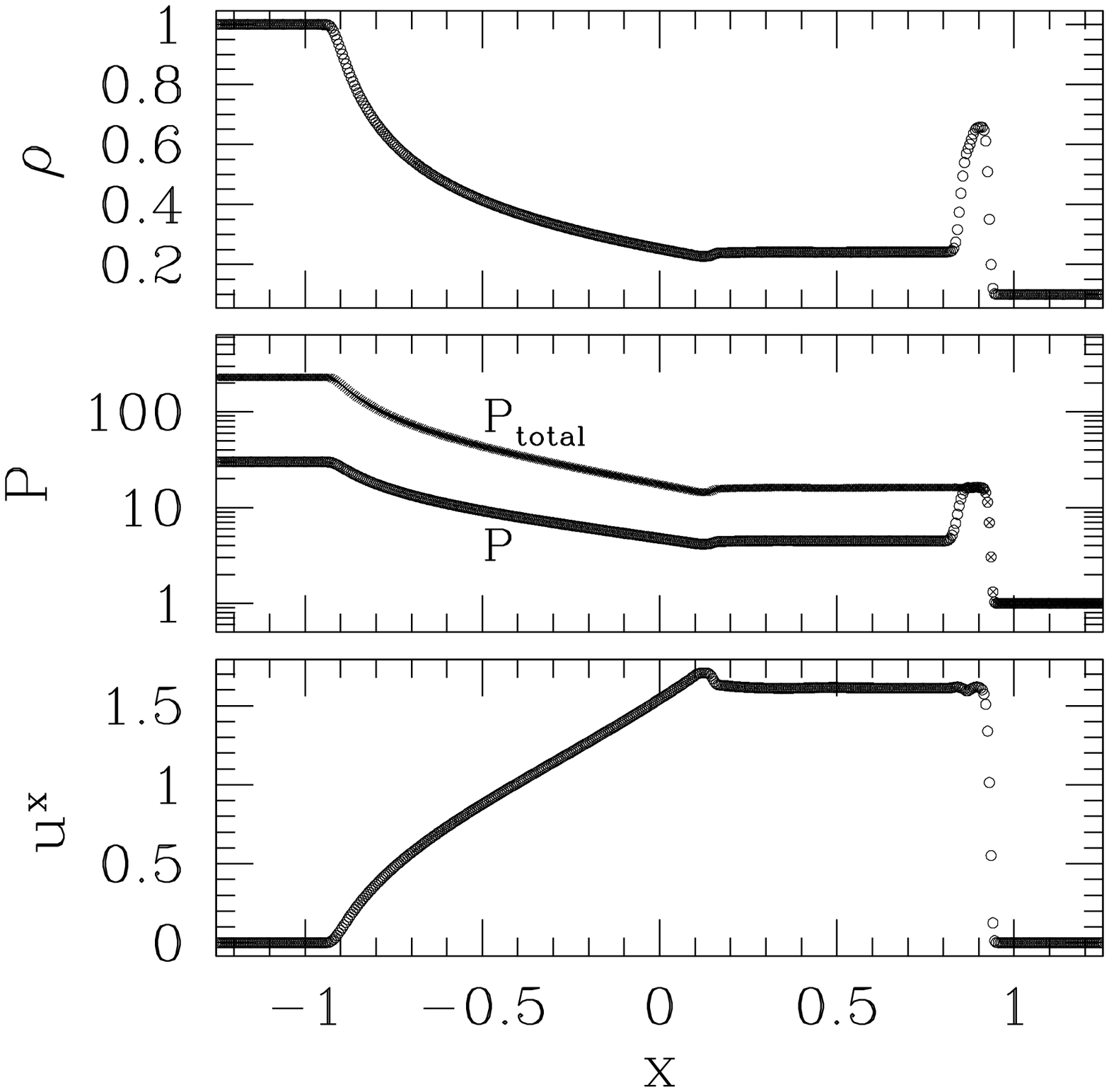} 
\end{center}
\vspace{-2mm}
\caption{(a) Shock tube problem with the initial discontinuities at
$x=0$ normal to the magnetic field. The 
numerical simulation is performed with $N=500$ and $\Delta x=0.005$.
We note that a large bump at $x \sim 0.9$ for $u^x$ is due to
a numerical error associated with the limiter ($b=2$) of a weak dissipation.
The height of this bump is decreased
if we use a more dissipative limiter (e.g.,
the minmod limiter with $b=1$). 
(b) Shock tube problem with the initial discontinuities at
$x=0$ parallel to the magnetic field.  The 
numerical simulation is performed with $N=500$ and $\Delta x=0.005$. 
For both cases, the snapshot at $t=1.0$ is shown. 
\label{FIG4} }
\end{figure}

\begin{figure}[thb]
\vspace{-4mm}
\begin{center}
\epsfxsize=3.in
\leavevmode
\epsffile{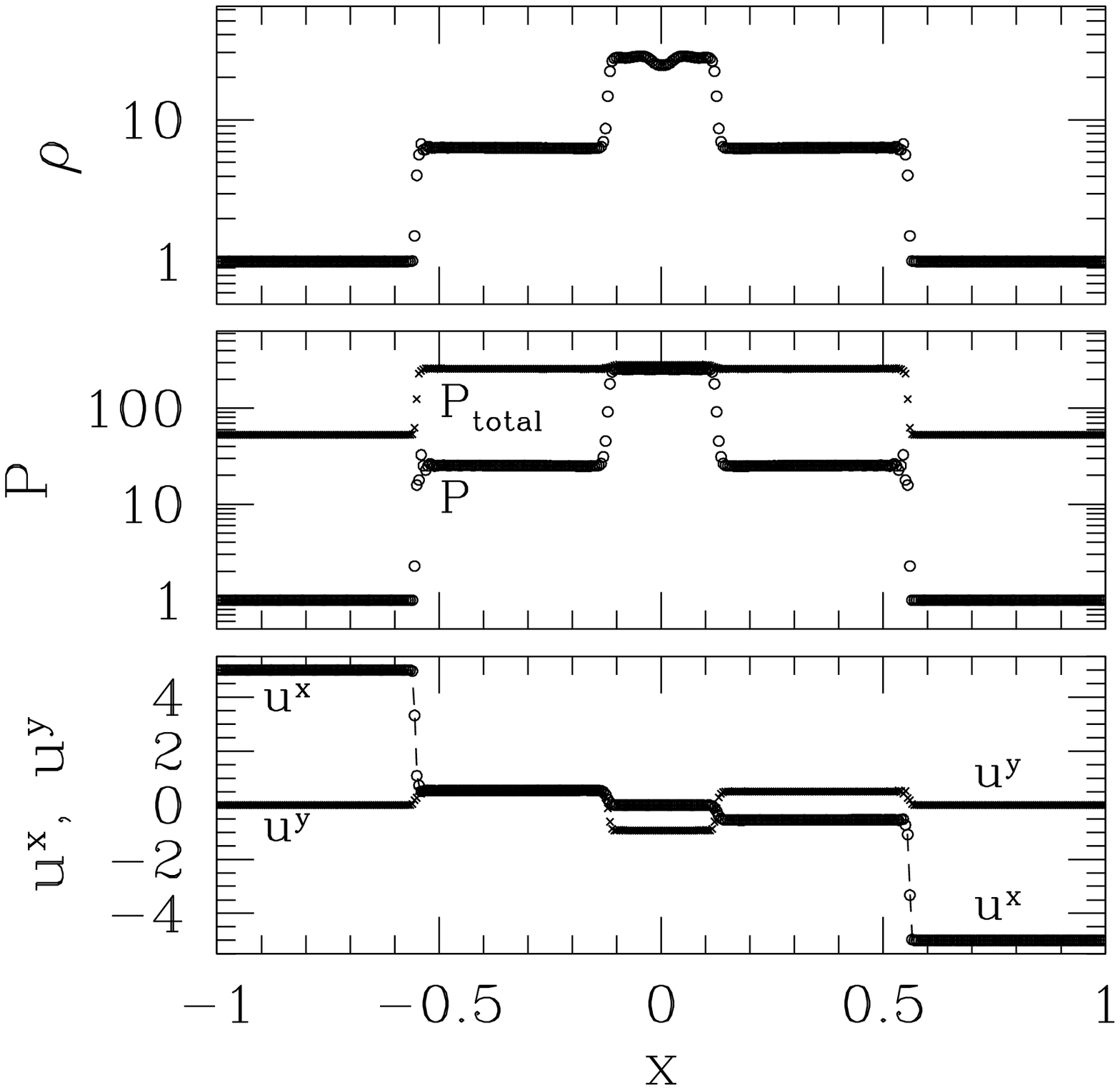} 
\end{center}
\vspace{-6mm}
\caption{Collision of two flows with opposite directions of
the tangential component of the magnetic field. 
The snapshot at $t=1.22$ is shown. The 
numerical simulation is performed with $N=400$ and $\Delta x=0.005$. 
\label{FIG5} }
\end{figure}

Any numerical implementation of the MHD equations
has to be checked if it can produce the basic waves such as
shock and rarefaction waves accurately.
Komissarov \cite{komissarov} has proposed
a suite of one-dimensional test problems in special relativity:
Propagation of fast and slow shocks, fast and slow rarefaction waves,
Alfv\'en waves, compound waves, shock tube tests, and collision of two
flows. We have performed all the tests except for the compound wave
following \cite{DLSS}. 
Our implementation can integrate each of remaining eight tests
although in some cases we have to reduce the Courant number significantly
to avoid numerical instabilities 
as reported by Gammie \cite{gammie}. On the other hand, we adopt
the same limiter, $b=2$, for all the simulations. 
Numerical results are shown in Figs. \ref{FIG1}--\ref{FIG5}.
Grid size $N$ and spacing $\Delta x$ we adopt for each of test
simulations are approximately the same as those by Komissarov, and 
described in the figure captions.

Figure \ref{FIG1} shows the results for
fast and slow shocks. In these problems, the system is stationary
with respect to the frame comoving with the shock front.
The velocity of the shocks is 0.2 and 0.5 for the
fast and slow shocks, respectively. As the previous works illustrate
\cite{komissarov,gammie,VH,DLSS}, the fast shock can be computed
accurately with a relatively large grid spacing. On the other hand,
in the numerical solution of the slow shock, a spurious
modulation is found for $\rho$ in the region of $1 \alt x \alt 1.3$ as in the
previous works \cite{komissarov,gammie,VH,DLSS}. 
This is always generated soon after the onset of the simulation
irrespective of grid resolutions. Thus, it is impossible to avoid
such small error in our implementation. Although the
modulation is always present, its wavelength and amplitude 
gradually decrease with improving the
grid resolution. We computed an L1 norm defined for the
difference between the numerical and exact solutions, and found that
it decreases as the grid spacing is smaller. In this case, 
the convergence is achieved at first order since discontinuities are present, 
around which the transport terms of hydrodynamic
equations are computed with the first-order accuracy. 

Figure \ref{FIG2} shows the results for switch-off and
switch-on rarefaction waves. Although we have not compared the results
precisely with those by other authors
\cite{komissarov,gammie,VH,DLSS}, the accuracy of our results is
similar to that reported by others.  For the switch-off waves, a
spurious bump is found at $x\sim -0.6$ as in the previous works 
\cite{komissarov,gammie,VH,DLSS}. As in the
slow shock problem, this bump is generated at $t=0$ irrespective of grid
resolutions, and with improving the grid resolution, the magnitude of
the L1 norm decreases at first order. On the other hand, the
numerical solution for the switch-on waves, spurious bumps are not
present, and with $\Delta x =0.005$, a good convergent result appears
to be obtained with our implementation.

Figure \ref{FIG3}(a) shows the results for an Alfv\'en wave test,
demonstrating that the Alfv\'en wave can be computed accurately with our
implementation as in \cite{komissarov,DLSS}.  In this problem, the
density and pressure should be unchanged.  In our results, this is
achieved within $\sim 1\%$ error for $\Delta x=0.0025.$ Since no
discontinuities are present in this problem, the convergence of the
numerical solution to the exact one should be achieved approximately
at second order \cite{DLSS}. To check if this is the case, we compute 
an L1 norm defined by the difference between the numerical and exact
solutions for $\rho$ and $P$.  The results are shown in
Fig. \ref{FIG3}(b), which illustrates that the convergence is
achieved approximately at second order (slightly better than second
order). 

In Fig. \ref{FIG4}, numerical results for shock-tube
problems are presented. For the problem of 
Fig. \ref{FIG4}(a), shocks are very strong since the 
ratio of the pressure in the left- and right-hand sides at $t=0$ is $10^3$. 
However, since the magnetic field lines are normal to the discontinuities, 
the effects of the magnetic field for the formation and
propagation of shocks are absent. In this case, 
a large spurious overshooting is found around the shock for $u^x$.
This is partly due to our limiter ($b=2$) which is not very
dissipative. If we use the minmod limiter ($b=1$), height of the 
overshooting decreases although the shocks are less sharply computed. 
For the problem of Fig. \ref{FIG4}(b),
shocks are not as strong as those in \ref{FIG4}(a).
However, the magnetic fields affect the formation and propagation of
shocks since they are parallel to the shocks.
The results shown in Fig. \ref{FIG4}(b) are very similar to
those in \cite{komissarov,gammie,VH,DLSS}, and hence, are
likely to be as accurate as them. This indicates that 
our implementation can compute magnetized shocks as
accurately as the previous ones. 

In Fig. \ref{FIG5}, numerical results for collision of two magnetized
flows are presented. It shows that four separate discontinuities
generated at $t=0$ are computed accurately. As found in previous papers
\cite{gammie,DLSS}, a small dip spuriously appears in $\rho$ around
$x=0$.  As in the case of the slow shock and switch-off rarefaction
wave, this is spuriously generated at $t=0$ irrespective of grid
resolution, and with improving the resolution, the magnitude of the
error is decreased at first order.

\subsection{Multi dimensional tests}

\begin{figure}[p]
\vspace{-4mm}
\begin{center}
\epsfxsize=4.in
\leavevmode
\epsffile{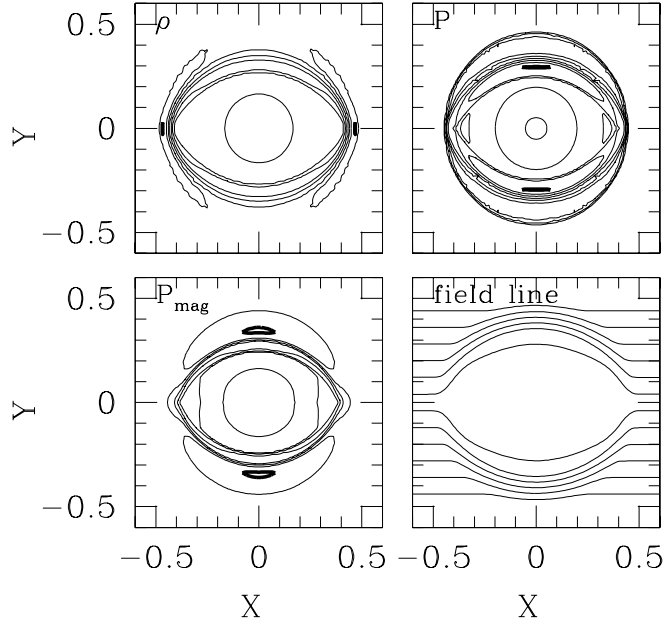} 
\end{center}
\vspace{-14mm}
\caption{Snapshot of cylindrical blast explosion (multidimensional
test (i)) at $t=0.4$. The contour curves for the 
density, pressure, and magnetic pressure $P_{\rm mag}$,
as well as the magnetic field lines are shown.
The contour curves for each quantity (denoted by $Q$) are drawn for
$Q=Q_{\rm max} \times 10^{-0.4\times i}$~$(i=1$--8) (solid curves)
and $Q=Q_{\rm max} \times 10^{-0.01}$ (thick solid curves).
Here $Q_{\rm max}$ denotes the maximum value. 
The results with $\Delta x= 0.004$ are presented. 
\label{FIG6} }
\end{figure}

\begin{figure}[p]
\vspace{-4mm}
\begin{center}
\epsfxsize=2.5in
\leavevmode
\epsffile{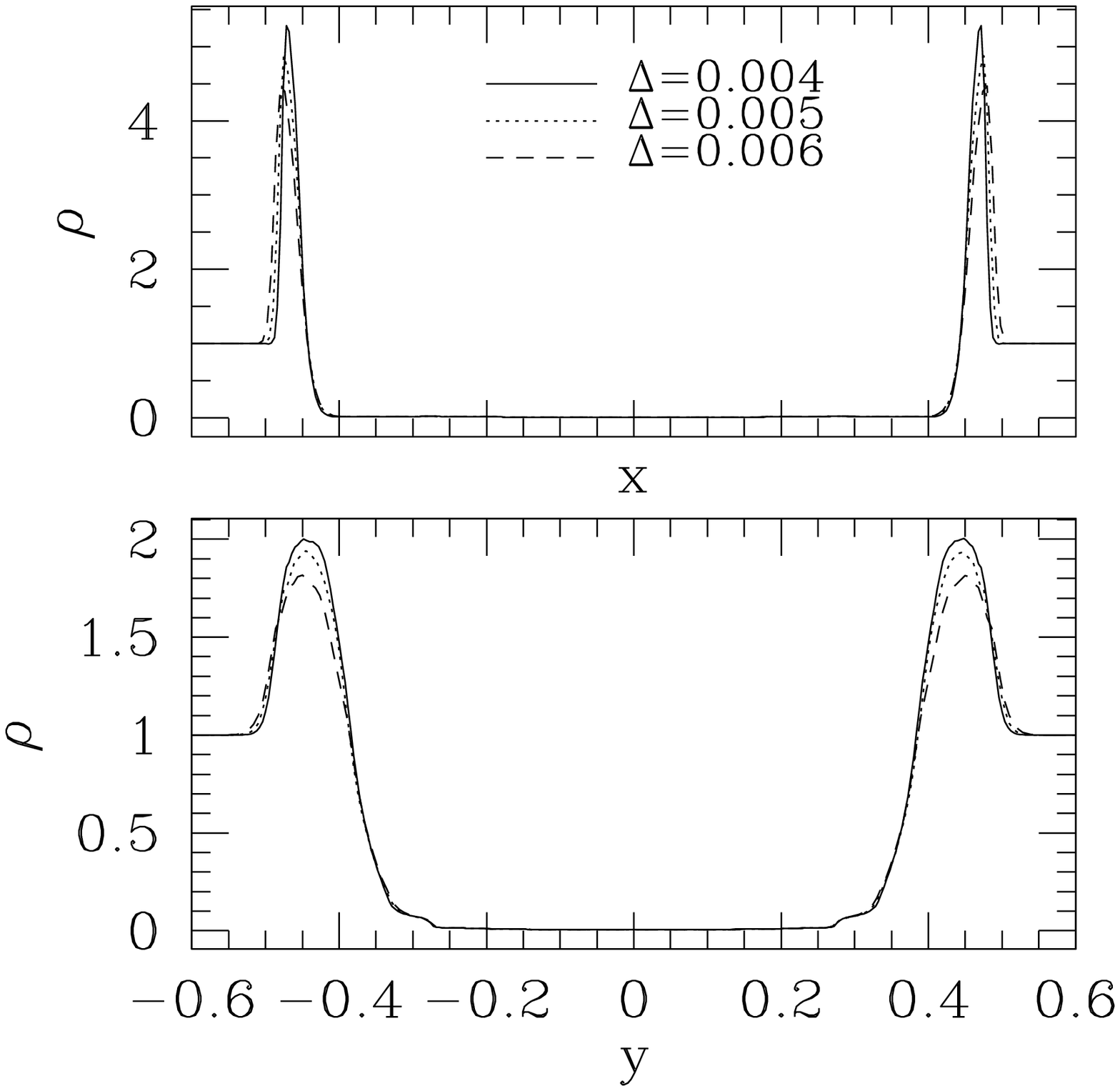} 
\epsfxsize=2.5in
\leavevmode
~~~\epsffile{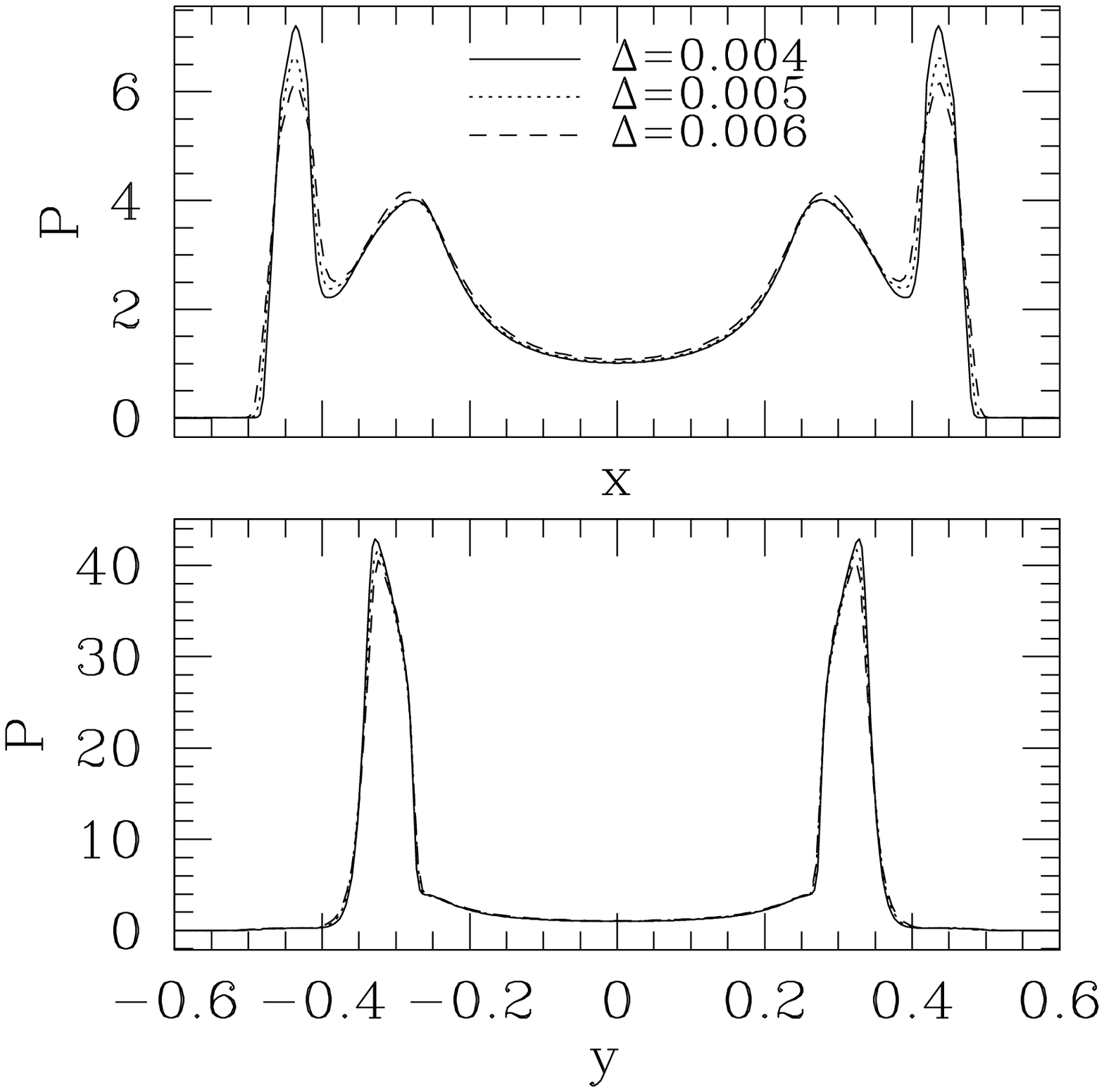}  \\
\vspace{-6mm}
\epsfxsize=2.5in
\leavevmode
\epsffile{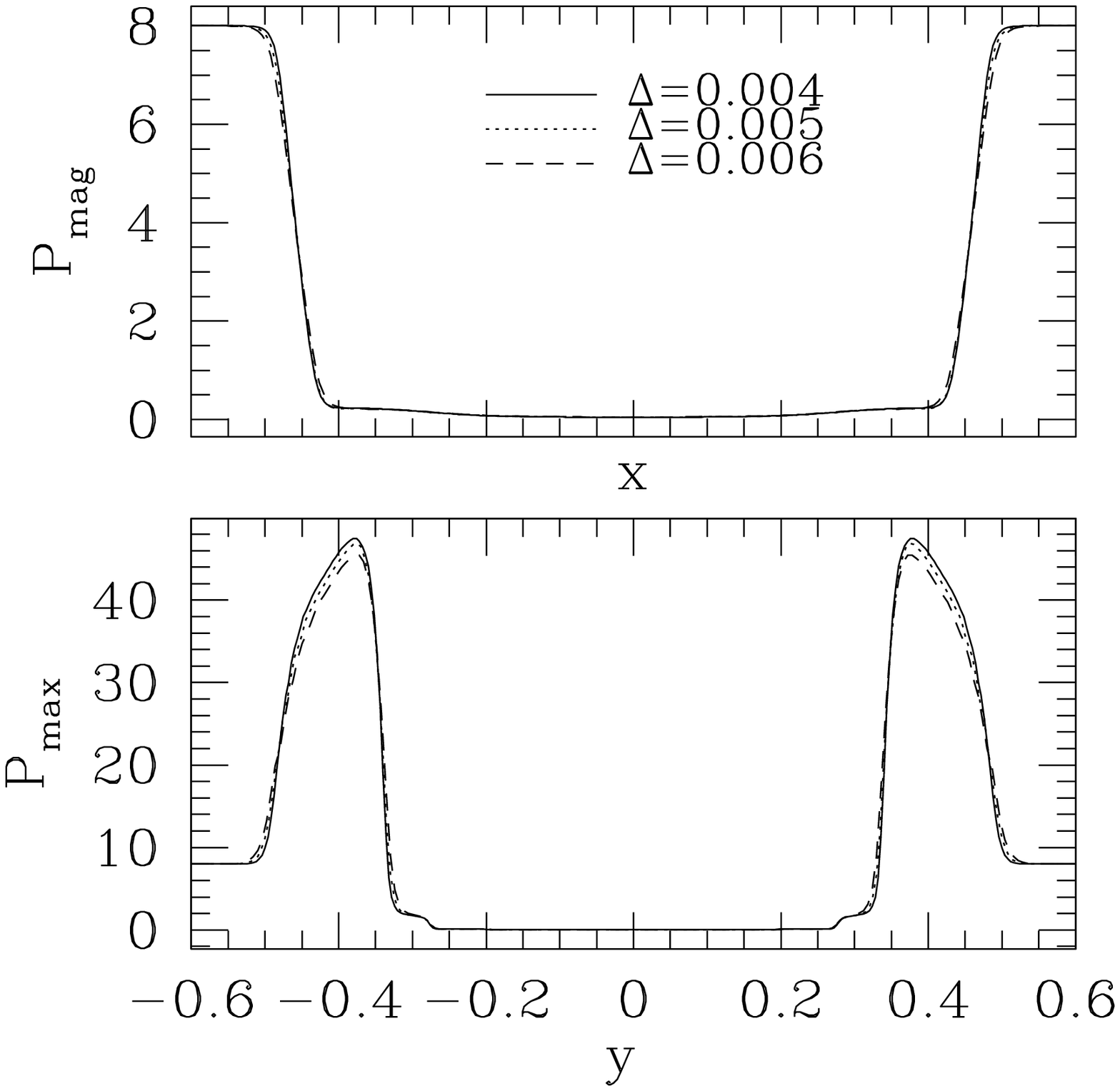} 
\epsfxsize=2.5in
\leavevmode
~~~\epsffile{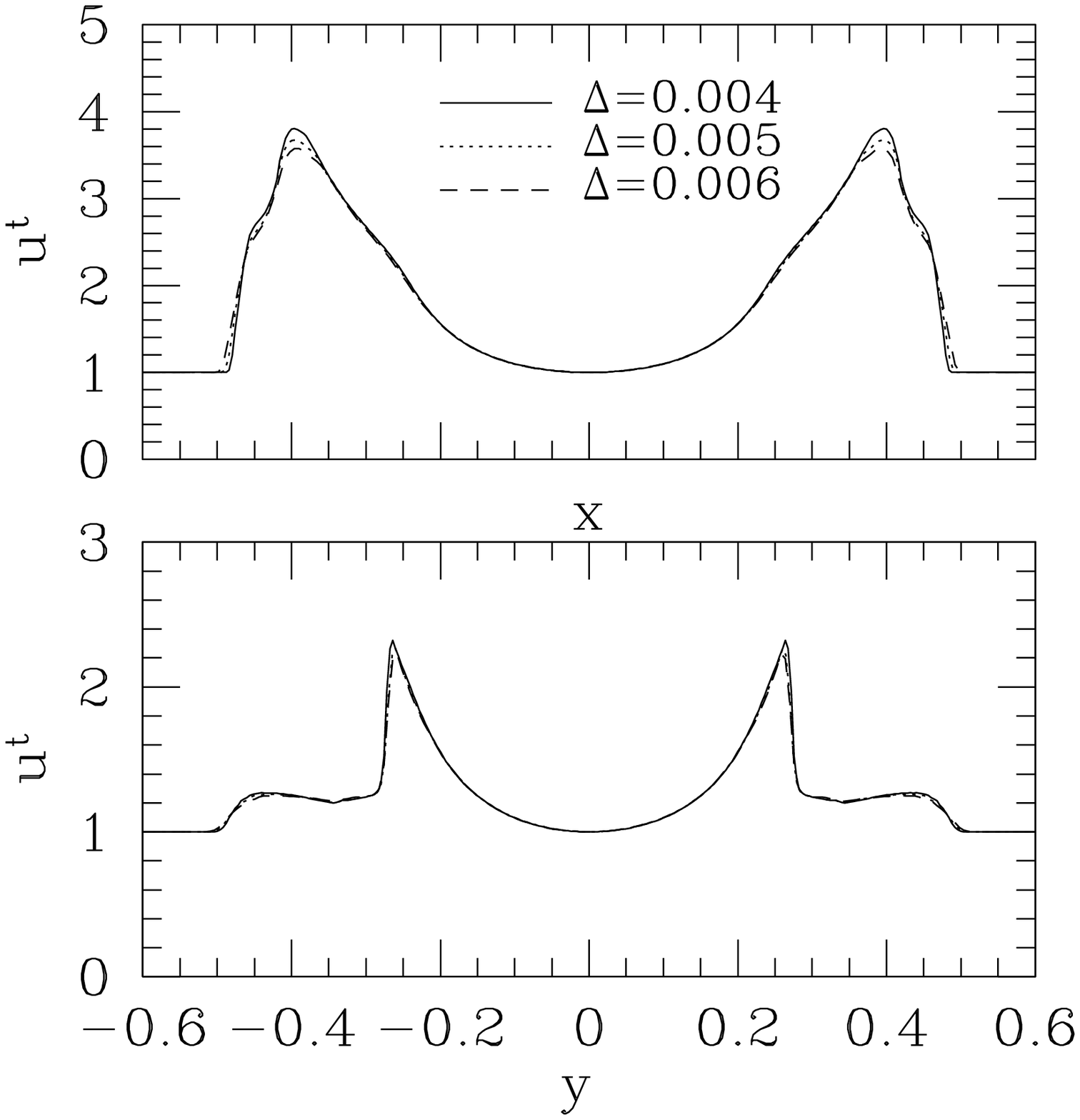} 
\end{center}
\vspace{-8mm}
\caption{Configuration of various quantities 
in cylindrical blast explosion along $x$ and $y$ axes at $t=0.4$ with 
different grid resolutions ($\Delta$ is the grid spacing). 
\label{FIG7} }
\end{figure}

\begin{figure}[p]
\vspace{-4mm}
\begin{center}
\epsfxsize=4.in
\leavevmode
\epsffile{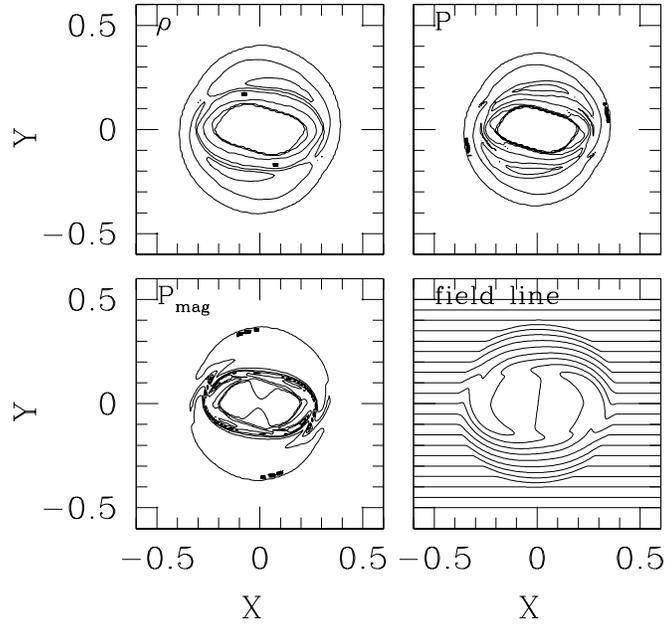} 
\end{center}
\vspace{-14mm}
\caption{The same as Fig. \ref{FIG6} but for an explosion of
a rotating cylinder (multidimensional test (ii)) at $t=0.4$.
The grid spacing for the corresponding simulation is 0.0025. 
\label{FIG8} }
\end{figure}

\begin{figure}[p]
\vspace{-4mm}
\begin{center}
\epsfxsize=2.5in
\leavevmode
\epsffile{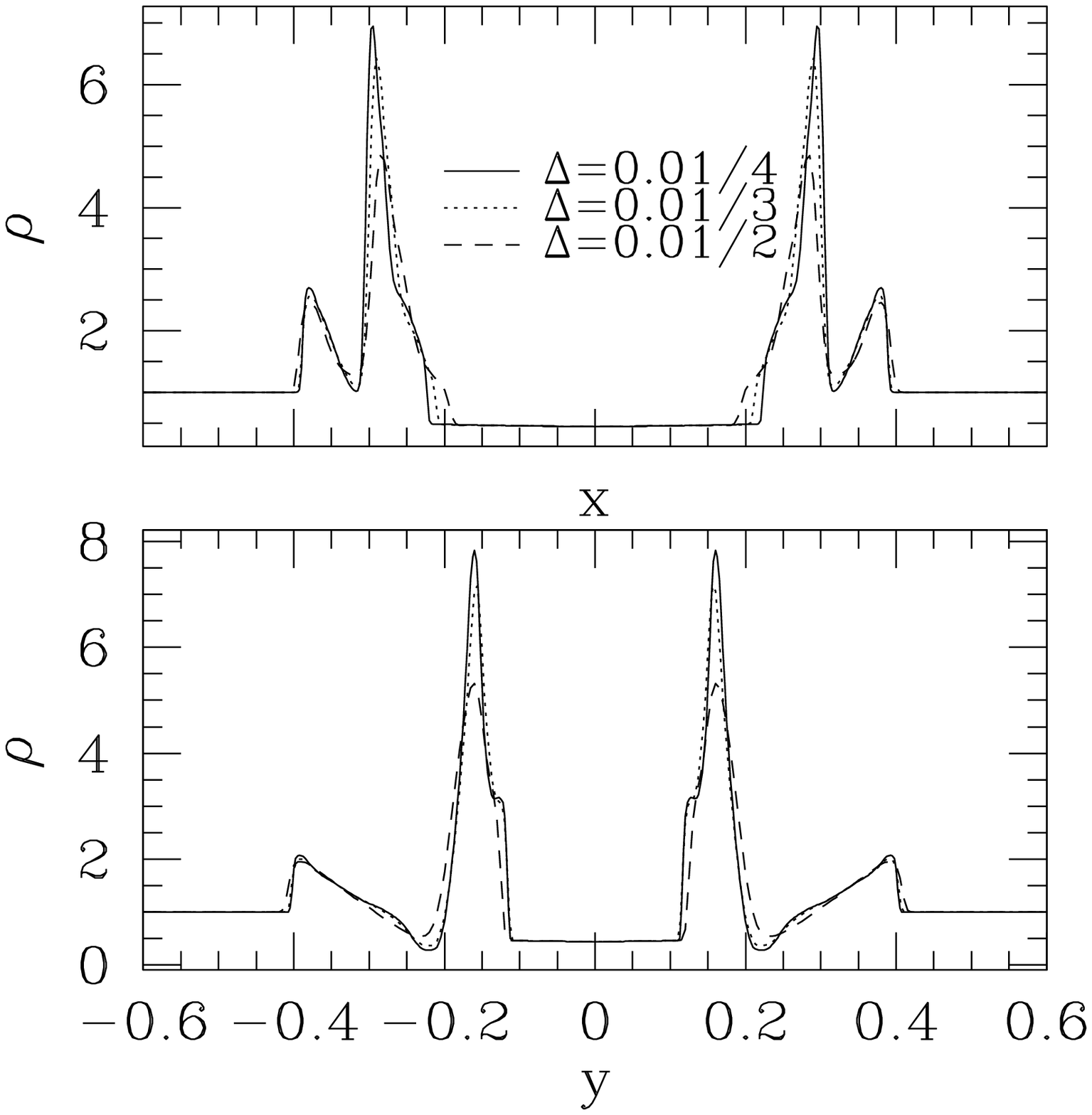} 
\epsfxsize=2.5in
\leavevmode
~~~\epsffile{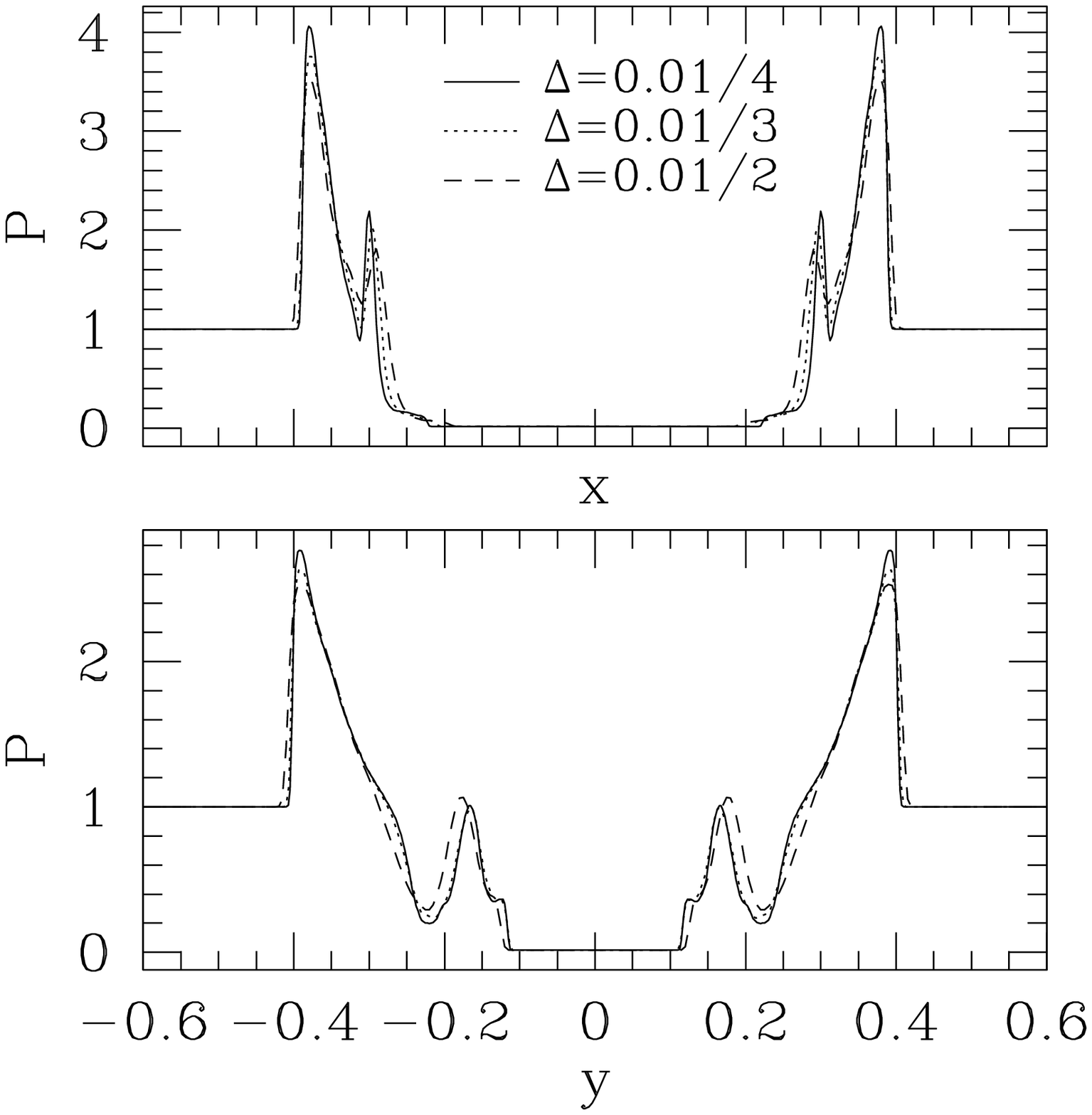}  \\
\vspace{-8mm}
\epsfxsize=2.5in
\leavevmode
\epsffile{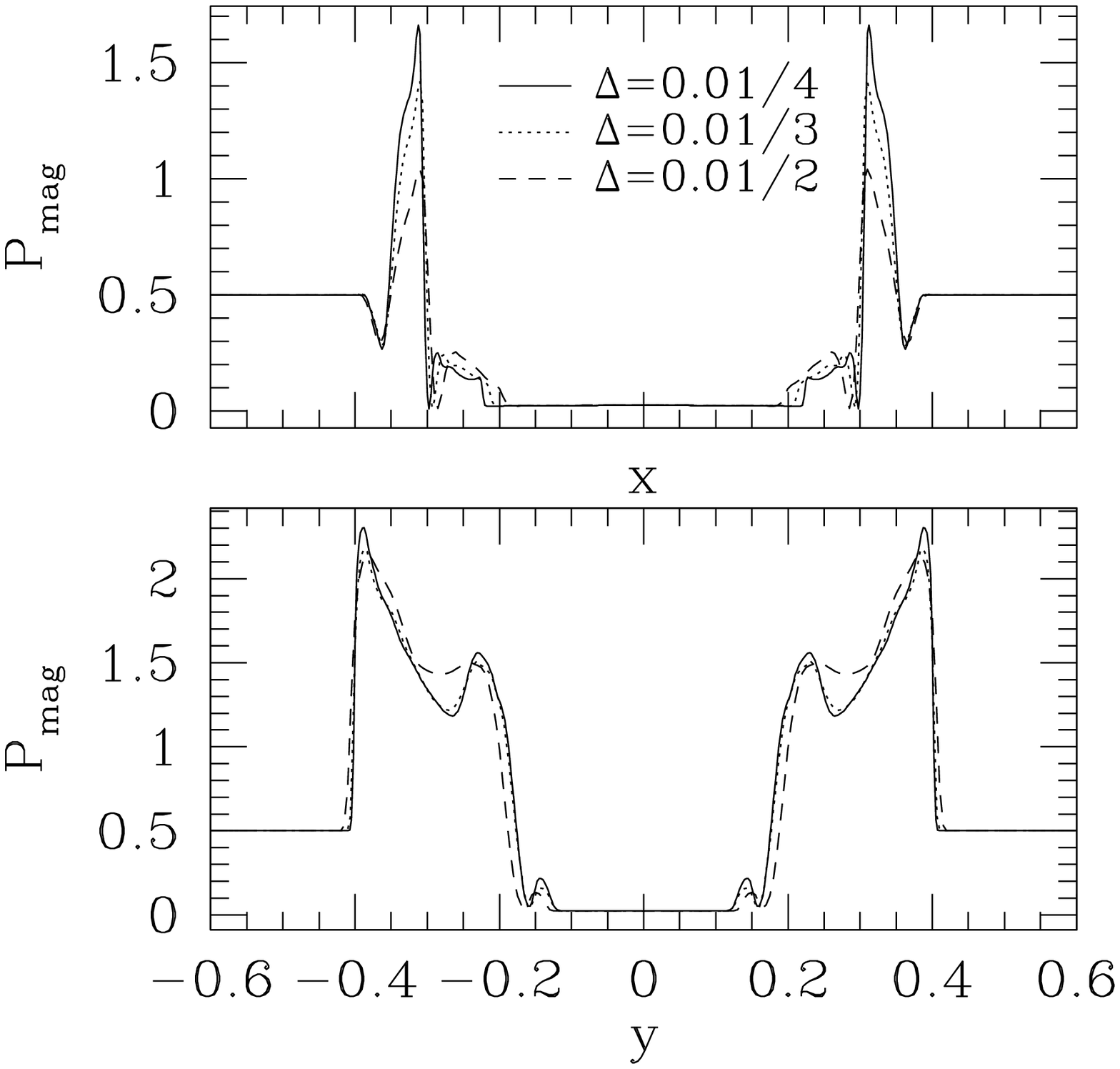} 
\epsfxsize=2.5in
\leavevmode
~~~\epsffile{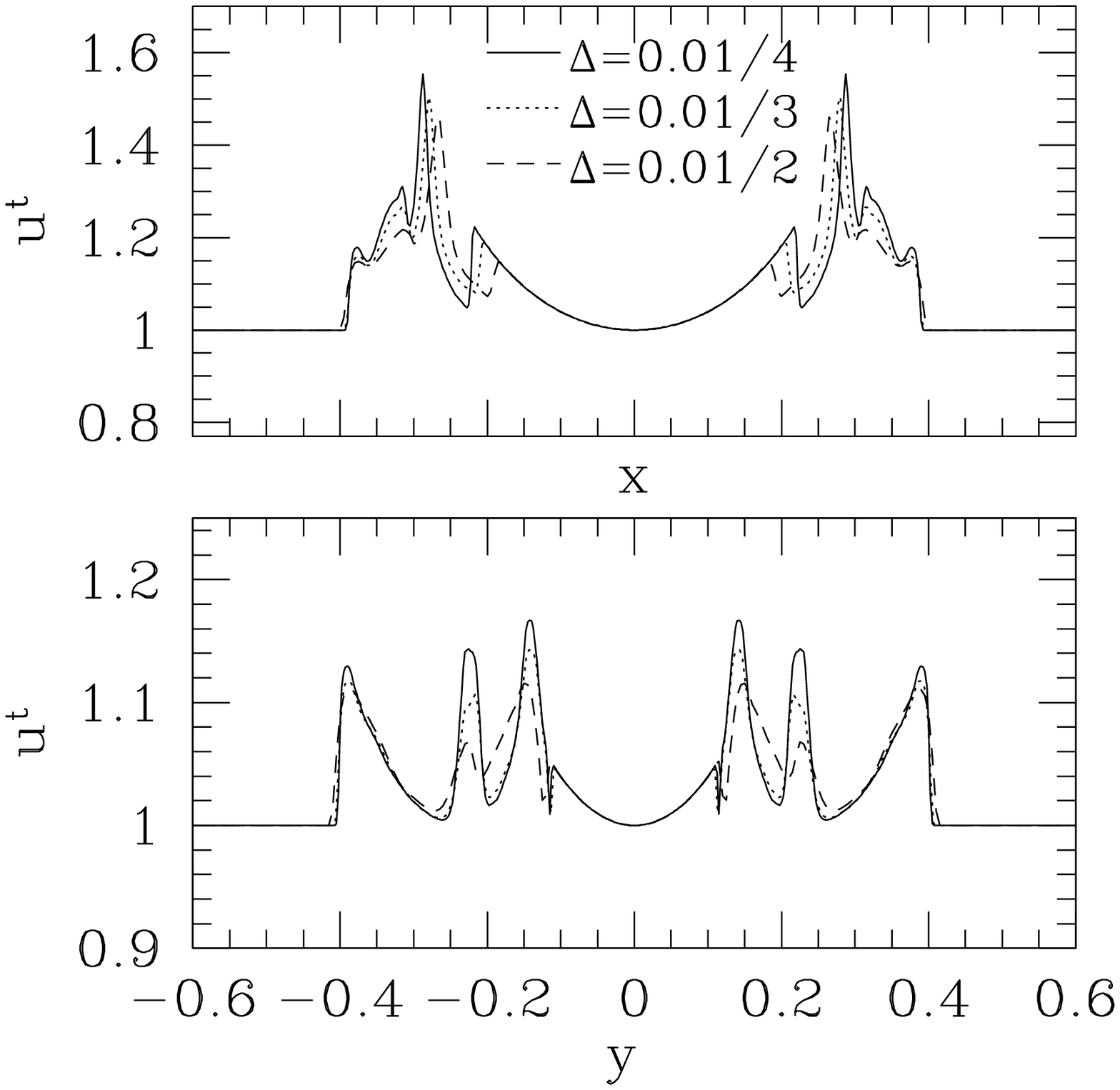} 
\end{center}
\vspace{-8mm}
\caption{Configuration of various quantities 
in an explosion of a rotating cylinder at $t=0.4$ with
different grid resolutions ($\Delta$ denotes the grid spacing). 
\label{FIG9} }
\end{figure}

\begin{figure}[thb]
\vspace{-4mm}
\begin{center}
\epsfxsize=3.7in
\leavevmode
\epsffile{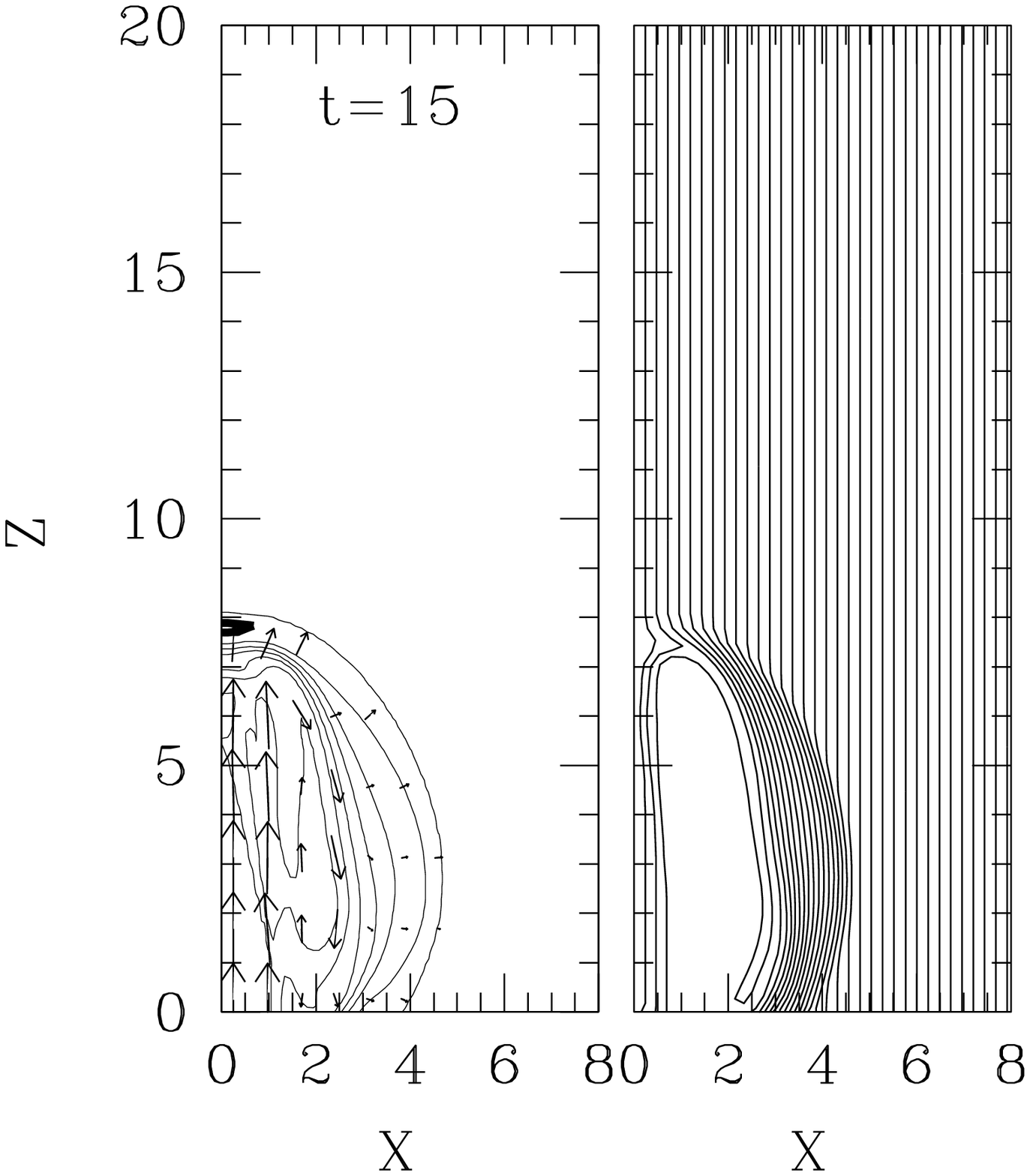} 
\hspace{-15mm}
\epsfxsize=3.7in
\leavevmode
\epsffile{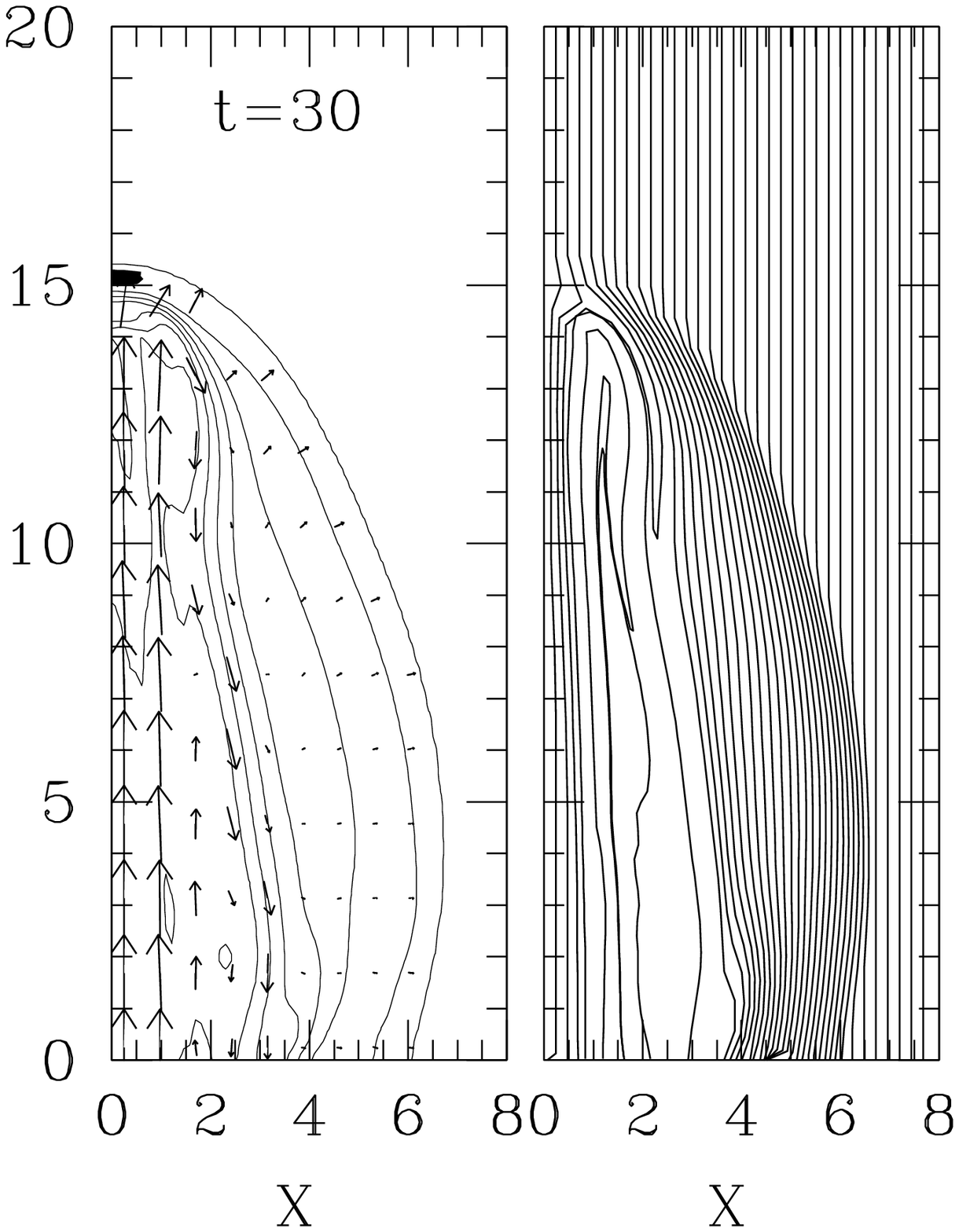} 
\end{center}
\vspace{-10mm}
\caption{The density contour curves and velocity vectors (left panel) and
the magnetic field lines (right panel) at $t=15$ and 
$t=30$ for an axisymmetric jet. 
The grid spacing for the corresponding simulation is 0.06.
The density contour curves are drawn by the same method as in
Fig. \ref{FIG6}. 
\label{FIG10} }
\end{figure}

For multidimensional tests, following Del Zanna et al. \cite{delzanna}, 
we performed simulations for (i) a cylindrical blast explosion,
(ii) a rotating cylinder in two-dimensional Cartesian
coordinates with a uniform magnetized medium, and (iii) propagation of a
jet in cylindrical coordinates in a magnetized background. 
The parameters for the initial conditions adopted here are the
same as those in \cite{delzanna}. On the other hand, we varied 
the grid spacing to see the convergence in contrast to the
previous works. 

In the test (i), the Cartesian grid of $(x, y)$ is adopted with the
range $[-0.6,0.6]$ for both directions. The grid spacing 
chosen is 0.004, 0.005, and 0.006. The initial condition is
\beqn
&& (\rho, P, \cB^x, \cB^y)=
\Big\{
\begin{array}{ll}
(1, 10^3, 4, 0)    &{\rm for}~~\sqrt{x^2+y^2} \leq 0.08, \\
(1, 10^{-2}, 4, 0) &{\rm for}~~\sqrt{x^2+y^2} > 0.08, 
\end{array}
\eeqn
with $u^i=0$ and $\Gamma=4/3$. Because of the large internal energy in
the central region, the outward explosion occurs. In this problem, the
shocks generated at $t=0$ are very strong, and hence, the minmod
limiter with $b=1$ is adopted to avoid numerical instability. With the
limiter of $b=2$, the computation crashes because of the appearance of
negative internal energy (or $h <1$) irrespective of the grid
resolution. We have found that the computation first crashes along the
line of $x=y$ and $x=-y$ for which the accuracy is likely to be worst.

In Fig. \ref{FIG6}, we display the snapshot of the numerical results
at $t=0.4$. In Fig. \ref{FIG7}, configurations of the density, pressure,
magnetic pressure, and Lorentz factor along $x$ and $y$ axes are
shown for three levels of grid resolution. The expansion velocity of the
blast wave is largest along the $x$ axis because of the confinement 
by the magnetic pressure. The maximum value of the Lorentz factor is
about $4$ at $t=0.4$ with the best resolved case. Along the $y$ axis,
the magnetic field lines are squeezed yielding the highest magnetic
pressure. These features agree with those found
in \cite{komissarov,delzanna}. As mentioned in \cite{delzanna}, the
total energy is completely conserved since we solve the MHD equations
in the conservative form and do not add any dissipative 
terms in contrast with the treatment in \cite{komissarov}.

One point to be mentioned is that convergence around the density peak
along the $x$ axis is not achieved well within the adopted resolution
although for other region, convergence is achieved well.
The likely reason is that the discontinuities around the peak is
very thin for which it is very difficult to resolve with the chosen
grid resolutions. Thus, it is difficult to accurately derive
the maximum values of the density, pressure, and Lorentz factor which
are underestimated in this test problem. 

In the test (ii), the Cartesian grid of $(x, y)$ is also adopted with the
range $[-0.6,0.6]$ for both directions. The grid spacing chosen is 
0.0025, 0.003, and 0.004. The initial condition is 
\beqn
&& (\rho, P, \cB^x, \cB^y)=
\Big\{
\begin{array}{ll}
(10, 1, 1, 0)    &{\rm for}~~\sqrt{x^2+y^2} \leq 0.1 \\
(1, 1, 1, 0) &{\rm for}~~\sqrt{x^2+y^2} > 0.1,
\end{array}
\eeqn
with $v^i=[-\omega y, \omega x]$ for $\sqrt{x^2+y^2} \leq 0.1$
where $\omega=0.995$
and thus the Lorentz factor at the surface of the rotating
cylinder is initially about 10. $\Gamma$ is chosen to be 5/3 
following \cite{delzanna}.

In Fig. \ref{FIG8}, we display the snapshot of the numerical results
at $t=0.4$. In Fig. \ref{FIG9}, configurations of the density, pressure,
magnetic pressure, and Lorentz factor along $x$ and $y$ axes are
shown for three levels of grid resolution. In this problem,
the magnetic field lines keep winding-up, and at $t=0.4$,
the central field lines are rotated by an angle of $\sim 90$ degrees.
Because of magnetic braking, the rotational speed is decreased
monotonically, and at $t=0.4$ the maximum Lorentz factor is decreased to
$\sim 1.7$. 
Due to the outward explosion induced by the rotation, the density in
the central region becomes an uniformly low value of
$\sim 0.44$ while an ellipsoidal
density peak is formed around the central region.
As in the test (i), it is difficult to 
obtain a convergent value for the peak density with the
chosen grid resolutions. The likely reason is that the thickness of
the density peak is so small that the grid resolutions are not sufficient.
However, for the other region, convergent results are obtained. 

In the test (iii), the cylindrical grid of $(x, z)$ is adopted with the
range $[0,8]$ and $[0,20]$, respectively. The grid spacing is
0.06, 0.08, and 0.1. The initial condition is 
\beqn
&& (\rho, P, v^z, \cB^z)=
\Big\{
\begin{array}{ll}
(10, 10^{-2}, 0.99, 0.1) &{\rm for}~~ 0 \leq x \leq 1~~{\rm and}~~
0 \leq z \leq 1 \\
(0.1, 10^{-2}, 0, 0.1)   &{\rm otherwise},
\end{array}
\eeqn
with $v^{x}=0$, $\cB^{x}=0$, and $\Gamma= 5/3$. The region with
$0 \leq x \leq 1$ and $0 \leq z \leq 1$ is defined to be a jet-inlet zone,
and the stationary condition is artificially imposed. In the simulation, 
the regularity condition is imposed  along the symmetric axis $x=0$.
For the boundary conditions at $z=0$, extrapolation is assumed following
\cite{delzanna}.
In this test, we adopt the minmod limiter with $b=1$ since with $b=2$,
the computation soon crashes irrespective of the grid resolutions. 

In Fig. \ref{FIG10}, we show the snapshot of the density contour
curves and magnetic field lines at $t=15$ and $30$ with $\Delta x=0.06$.
The contour curves and field lines are similar to those in \cite{delzanna}.
The maximum Lorentz factor at $t=0$ is $\sim 7.09$.
At the head of the jet, the density becomes maximum and
shocks are formed, inducing back flows at the shocks.
These flows make a cocoon which
is to expand in the direction of the cylindrical
radius, squeezing the magnetic field lines. A part of the 
matter is back-scattered toward the $z=0$ plane dragging 
the magnetic field lines together. As a result, the
magnetic field lines are highly deformed.
The deformation is computed more accurately with finer grid resolutions. 

However, we found that precise computation for the deformation of the
magnetic field lines increases the risk for crash of the
computation. For $\Delta x=0.1$ and 0.08, computations can be
continued until the shock front of the jets reaches the outer
boundary. However, for $\Delta x=0.06$, the computation crashes at $t
\sim 35$ in spite of the fact that the motion of the jet head is still
stably computed. If we adopt a better resolution with $\Delta x \alt
0.05$, the computation crashes before $t$ reaches 30. The
instabilities always occur near the boundary region of the jet-inlet
zone around which the magnetic field configuration is deformed to be
highly complicated. This seems to be due to the fact that we impose
the stationary condition inside the jet-inlet zone. This artificial
handling makes the field configuration near the boundary of the
jet-inlet zone nonsmooth (i.e., the derivative of the magnetic field
variables can be artificially larger for
better grid resolutions). Here, we note that this problem happens
only in the presence of magnetic fields. Thus, 
for continuing the computation for a longer 
time, probably, it is necessary to include a resistivity for inducing
reconnections of magnetic fields near the jet-inlet zone for 
stabilization. The other method is to change the stationary
condition we adopt here to other appropriate boundary conditions near the
jet-inlet zone \cite{Aloy2}.

\section{General relativistic tests}

\subsection{Relativistic Bondi accretion}

As the first test for general relativistic implementation, we perform
a simulation for spherical accretion onto the fixed background of
a Schwarzschild black hole. 
The relativistic Bondi solution is known to
describe a stationary flow, and thus, by comparing the numerical
solution with the analytical one, it is possible to check the suitability 
of the numerical implementation for general relativistic hydrodynamics
problems 
\cite{Bondicheck,SP}.  Furthermore, it has been shown that the relativistic
Bondi solution is unchanged even in the presence of a divergence-free
pure radial magnetic field \cite{VH}.  Thus, it can be also used for
checking the GRMHD implementations. The advantage
of this test is that the exact solution can be obtained very easily
while it involves strong gravitational fields, relativistic flows, and
strong magnetic fields all together.

Following previous authors \cite{gammie,DLSS}, we write the metric in
Kerr-Schild coordinates in which all the variables are well behaved at
the event horizon ($r=2M$; where $r$ and $M$ are the radial coordinate
and the mass of the black hole). Nevertheless, the hydrostatic
equations for the stationary flow are the same forms as those in the
Schwarzschild coordinates, and thus, the stationary solution is
determined from an algebraic equation which can be easily
solved by standard numerical methods \cite{ST}.

For this test, we adopt the same solution used in
\cite{Bondicheck,gammie,VH,DLSS}. Namely, the sonic radius is set at
$r=8M$, the accretion rate $\dot M=4\pi\rho r^2 u^r$ is set to be
$-1$, and the adiabatic index for the equation of state is $4/3$. The
simulation is performed in an axisymmetric implementation with the
cylindrical coordinates $(x, z)$. The computational domain is set to
be $[0, 18M]$ for $x$ and $z$, and the radius of $r=1.9M$ is chosen as
the excision radius. The uniform grid is adopted. At the excision
radius and outer boundaries, we impose the condition that the system
is stationary. The (semi) analytic solution for the stationary Bondi
flow is put as the initial condition, and we evolve for $100M$
following previous authors \cite{gammie,VH,DLSS}.

The simulations are performed changing the grid spacing $\Delta
x$. Irrespective of the grid resolution, the system relaxes to a
stationary state long before $100M$. When evolved with a
finite-difference implementation, discretization errors will cause
small deviations in the flow from the exact stationary configuration.
These deviations should converge to zero at second order with
improving the grid resolution. To diagnose the behavior of our
numerical solution, we measure an L1 norm for $\rho_*-\rho_*^{\rm
exact}$ where $\rho_*^{\rm exact}$ denotes the exact stationary value
of $\rho_*$.  Specifically, the L1 norm is here defined by
\beqn
\int_{r \geq 2M} |\rho_*-\rho_*^{\rm exact}| d^3x
\Big/ \int_{r \geq 2M} \rho_*^{\rm exact} d^3x. 
\eeqn
For the convergence test, the grid spacing is changed from
$0.06M$ to $0.4M$. 

The radial magnetic field strength is also changed for
a wide range. Following \cite{gammie,DLSS}, we denote the 
magnetic field strength by
\beqn
\hat \beta \equiv {b^2 \over \rho}\Big|_{r=2M}. \label{hatbeta}
\eeqn
We note that the ratio of the magnetic pressure to the gas pressure
$b^2/2P$ is $\approx 3.85\hat \beta$ at $r=2M$ for the solution chosen
in this test problem. 

\begin{figure}[thb]
\vspace{-4mm}
\begin{center}
\epsfxsize=4.in
\leavevmode
\epsffile{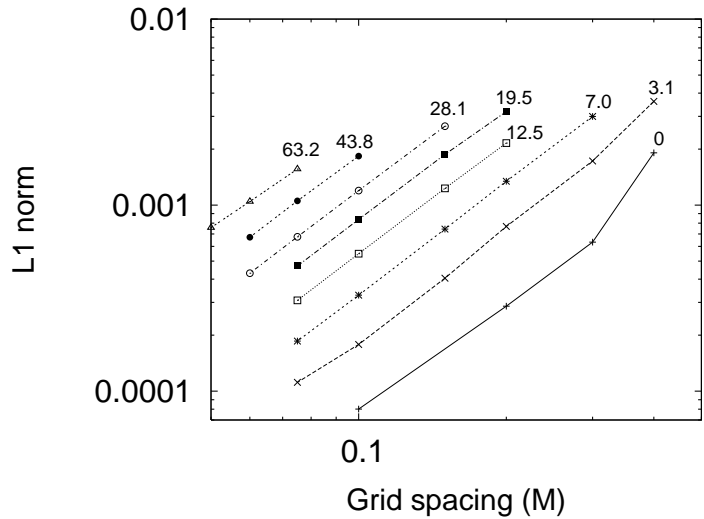} 
\end{center}
\caption{L1 norm for the error of density $\rho_*$
as a function of the grid spacing 
in units of $M$ for the magnetized Bondi flow. The numerical
numbers attached for each curve denote the values of $\hat \beta$
(see Eq. (\ref{hatbeta}) for its definition). 
\label{FIG11} }
\end{figure}

In Fig. \ref{FIG11}, we show the L1 norm as a function of the grid
spacing for $0 \leq \hat \beta \leq 63$. Irrespective of the magnetic
field strength, the numerical solution converges approximately at
second order to the exact solution for $\Delta x \rightarrow 0$. The
L1 norm is larger for the stronger magnetic fields, implying that the
relaxed state deviates more from the true stationary solution for the
larger value of $\hat \beta$. Specifically, the velocity field
configuration deviates significantly from the exact solution with 
increasing the value of $\hat \beta$, although the deviation for the
density configuration is not very outstanding.

In this test simulation, we have found several interesting behaviors of
the numerical solutions. 
First, for a given value of the grid spacing with $\Delta x \agt 0.1M$,
there is the maximum allowed value of $\hat \beta$ above which
the computation crashes. The maximum value is larger for better
grid resolution; e.g. for $\Delta x=0.1M$, $0.2M$, and $0.3M$,
the maximum allowed values of $\hat \beta$ are $\approx 45$, 
$25$, and $10$, respectively. For $\Delta x \alt 0.1M$, on the other hand,
the maximum allowed value is 
$\hat \beta \sim 70$ irrespective of the grid resolution. 
The limitation is due to the well-known weak point in the conservative
scheme that 
the small error in the magnetic energy density in the magnetically dominated
region with $\hat \beta \gg 1$ leads to fractionally large errors in other 
components of the total energy density, by which the computation crashes 
(typically, the internal energy density becomes negative). 
For the poorer grid resolutions, the numerical error is larger, and hence, 
the computation crashes for the smaller value of the magnetic field. 
Second, the maximum allowed value of $\hat \beta$
found here $(\sim 70)$ is by about one order of magnitude smaller than
that found in \cite{gammie}. This is probably due to the
difference of the coordinate system adopted; we use the cylindrical
coordinates while the authors in \cite{gammie} use the spherical
polar coordinates which obviously have advantage for handling the
spherically symmetric problem. However, we note that
even in the cylindrical coordinates, it is possible to
handle the flow with a very high value of $\hat \beta \sim 60$
if a sufficient grid resolution is guaranteed. In \cite{DLSS}, the
authors suggest that in the cylindrical coordinates, the
maximum allowed value of $\hat \beta$ is at most $\sim 10$. 
We have not found such severe limitation in our numerical experiment. 
Their failure for simulating the flow with high values of $\hat \beta$
is probably due to the fact that they use an 
excision boundary which may be applicable for more general problems
(e.g., for simulation of dynamical spacetimes).
Even in the cylindrical coordinates, a high value of $\beta$ will
be achieved if a stationary inner boundary condition is imposed. 

\subsection{Longterm evolution for system of a rotating star
and a disk with no magnetic field}

Next, we illustrate that with our implementation (for axisymmetric
systems), self-gravitating
objects can be simulated accurately. In a previous paper \cite{ShibaFont},
we have already illustrated that our implementation with a HRC scheme
can simulate rapidly rotating compact neutron stars for more than 20 
rotational periods accurately. Thus, we here choose 
a more complicated system; an equilibrium system 
composed of a rapidly rotating neutron star and a massive disk. 
By this test, it is possible to check that 
our implementation is applicable to a longterm evolution not only for
an isolated rotating star but also for a self-gravitating disk
rotating around a compact object.

The equilibrium configuration is determined by solving equations for
the gravitational field and hydrostatic equations self 
consistently. For simplicity, we here adopt a conformally flat 
formalism for the spatial metric \cite{WM}. As shown in \cite{CST}, a 
good approximate solution for axisymmetric rotating stars can be 
obtained even in this approximation.  Thus, the initial condition 
presented here can be regarded as a slightly perturbed equilibrium 
state. At the start of the simulations, we further add a slight 
perturbation by reducing the pressure by 0.1\% to investigate if a 
quasiradial oscillation is followed stably and accurately.  The
magnitude of the perturbation in association with the
conformally flat approximation is much smaller than this pressure
perturbation.

The Euler equation for axisymmetric stars in equilibrium can be
integrated to give the first integral, which is written as
\beqn
\ln {h \over u^t}+ \int u^t u_{\varphi} d\Omega =C ,\label{HS1}
\eeqn
or 
\beqn
{h \over u^t}+ \int h u_{\varphi} d\Omega =C', \label{HS2}
\eeqn
where $C$ and $C'$ are integral constants. Equation (\ref{HS1}) is
a well-known form \cite{Ster}. However here, we adopt Eq. (\ref{HS2}), and  
set that the specific angular momentum $h u_{\varphi}$ is constant $(=j)$
for the disk and $\Omega=$const for the central star. 

A hybrid, parametric equation of state is used in this simulation following
previous papers \cite{HD,SS,SS3d}. In this equation of state,
one assumes that the pressure consists of the sum of polytropic and
thermal parts as
\beq
P=P_{\rm P}+P_{\rm th}. \label{EOSII}
\eeq
The polytropic part, which denotes the cold part of
the equations of state, is given by 
\beqn
P_{\rm P}=
\left\{
\begin{array}{ll}
K_1 \rho^{\Gamma_1}, & \rho \leq \rho_{\rm nuc}, \\
K_2 \rho^{\Gamma_2}, & \rho \geq \rho_{\rm nuc}, \\
\end{array}
\right.\label{P12EOS}
\eeqn
where $K_1$ and $K_2$ are polytropic constants.  $\rho_{\rm nuc}$
denotes the nuclear density and is set to be $2\times 10^{14}~{\rm
g/cm^3}$. In this paper, we choose $\Gamma_1=4/3$ and $\Gamma_2=2.5$.
Since $P_{\rm P}$ should be continuous, the relation,
$K_2=K_1\rho_{\rm nuc}^{\Gamma_1-\Gamma_2}$, is required. 
Here, the value of $K_1$ is chosen to be $2.5534 \times 10^{14}$ in
the cgs unit. With this value, the maximum ADM (baryon rest) mass for
the cold and spherical neutron star becomes about $1.84M_{\odot}$
($2.05M_{\odot}$) which is a close to that derived in
realistic equations of state \cite{EOSEOS}.

Since the specific internal energy should be also continuous
at $\rho=\rho_{\rm nuc}$, the polytropic specific 
internal energy $\varepsilon_{\rm P}$ is defined as 
\beqn
\varepsilon_{\rm P}=
\left\{
\begin{array}{ll}
\displaystyle
{K_1 \over \Gamma_1-1} \rho^{\Gamma_1-1}, & \rho \leq \rho_{\rm nuc}, \\
\displaystyle 
{K_2 \over \Gamma_2-1} \rho^{\Gamma_2-1}
+{(\Gamma_2-\Gamma_1)K_1 \rho_{\rm nuc}^{\Gamma_1-1}
\over (\Gamma_1-1)(\Gamma_2-1)},  & \rho \geq \rho_{\rm nuc}. \\
\end{array}
\right.
\eeqn
The thermal part of the pressure $P_{\rm th}$ plays an important
role in the case 
that shocks are generated. $P_{\rm th}$ is related to the thermal energy
density $\varepsilon_{\rm th}\equiv \varepsilon-\varepsilon_{\rm P}$ as 
\beq
P_{\rm th}=(\Gamma_{\rm th}-1)\rho \varepsilon_{\rm th}. 
\eeq
For simplicity, the value of $\Gamma_{\rm th}$, which determines
the strength of shocks, is chosen to be equal to $\Gamma_1$.
For computing initial equilibria, we set $\varep=\varep_{\rm P}$
and $P=P_{\rm P}$. 

\begin{figure}[thb]
\vspace{-4mm}
\begin{center}
\epsfxsize=3.in
\leavevmode
(a)\epsffile{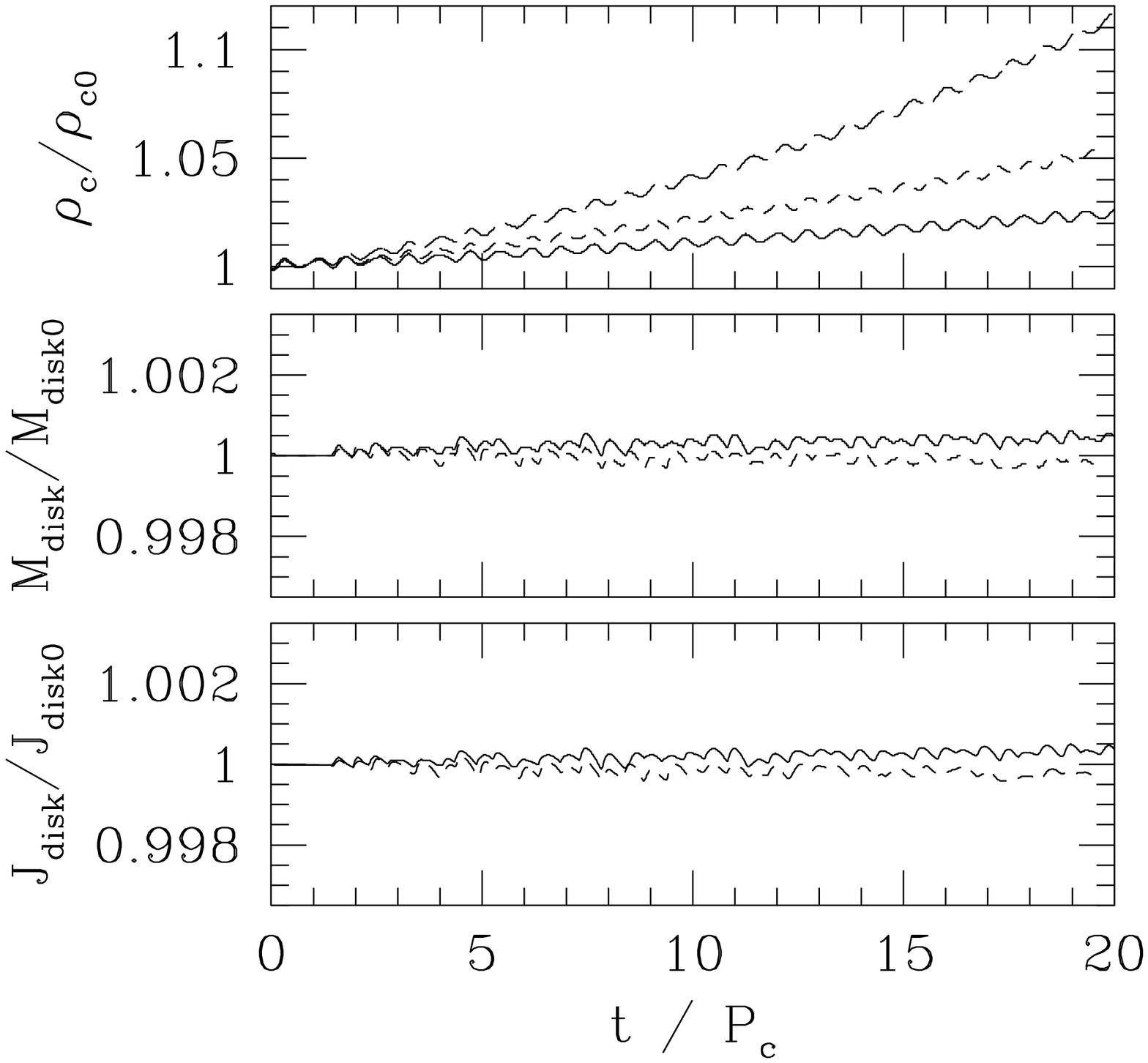} 
\epsfxsize=3.in
\leavevmode
~~~(b)\epsffile{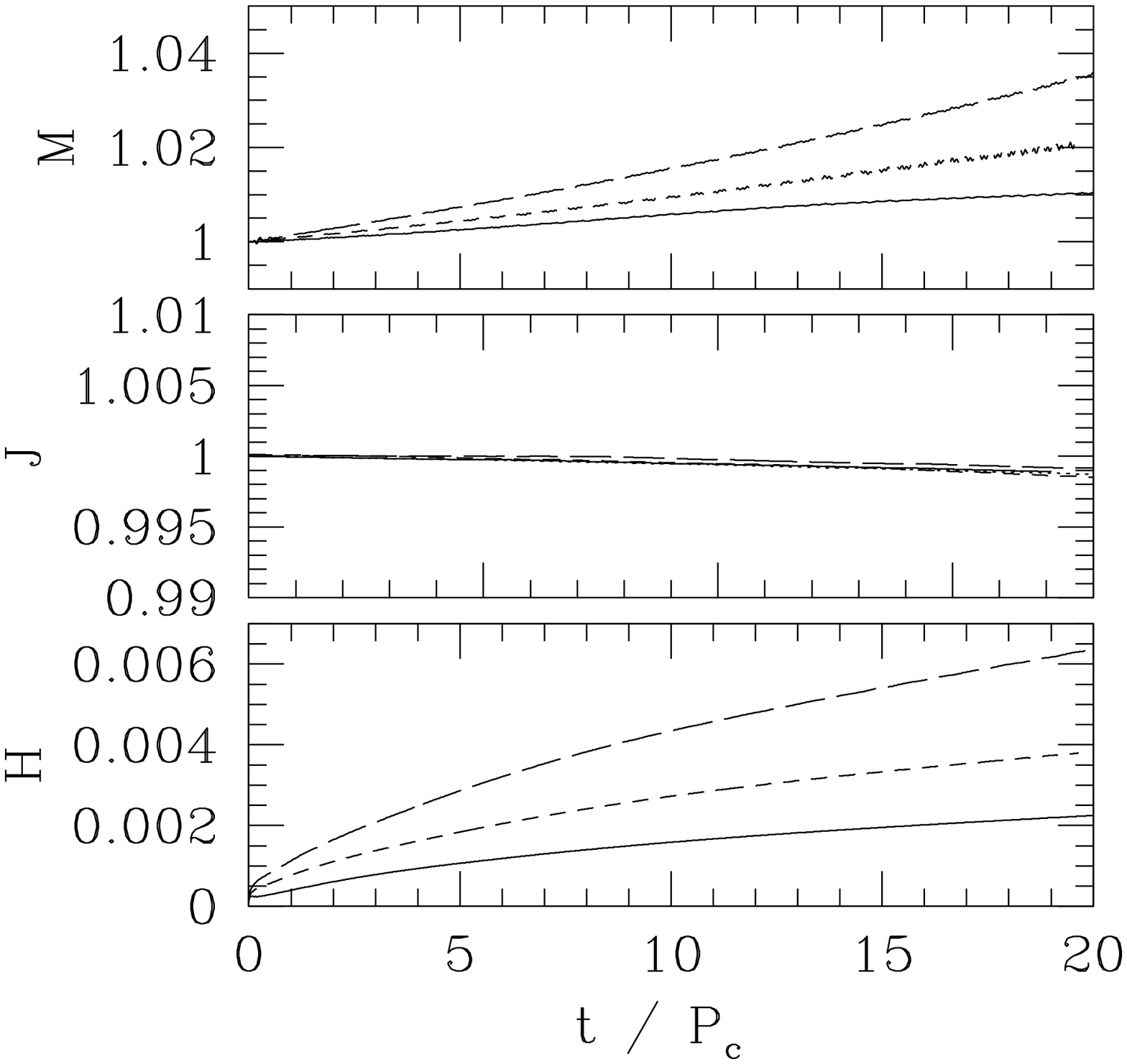} 
\end{center}
\vspace{-3mm}
\caption{(a) Evolution of the central density of a neutron star
and mass and angular momentum of a disk around the neutron
star. (b) Evolution of the ADM mass, angular momentum, and
averaged violation of the Hamiltonian constraint. The ADM mass and
angular momentum are shown in units of their initial values. 
The solid, dashed, and long-dashed curves show the results with 
(241,241), (193, 193), and (161,161) grid sizes, respectively. 
\label{FIG12} }
\end{figure}

For the simulation, we choose a sufficiently deformed star with the
axial ratio of the minor axis to major axis $\sim 0.6$. The ADM mass
is $1.888 M_{\odot}$, total baryon rest mass $2.074M_{\odot}$, the
central density $1.3 \times 10^{15}~{\rm g/cm^3}$, the circumferential
radius at equator $16.2$ km, the rotational period $P_c=0.841$
ms, and $J/M^2=0.545$. Thus, the neutron star considered is massive
and rapidly rotating.  The baryon rest mass of the disk is much
smaller than the central star as $4.9 \times 10^{-5} M_{\odot}$ with
the maximum density $\approx 2 \times 10^{10}~{\rm g/cm^3}$.  Since it
is of low density, the disk is composed of $\Gamma=4/3$ polytropic
equation of state.  Orbital radius of inner edges of the disk is $\sim
20$ km, and thus, the uniform specific angular momentum is small as $j
\approx 3.45M$ which is very close to the value for a particle
orbiting an innermost stable circular orbit. 
The rotational periods of the disk at the inner and
outer edges in the equatorial plane are 1.03 ms($=1.2 P_c$) and 2.58
ms($=3.1P_c$), respectively.  The simulations are performed in axial
symmetry with (241,241), (193, 193), and (161,161) grid sizes for
which the grid spacing is 0.165, 0.202, and 0.248 km,
respectively. The reflection symmetry with respect to the
equatorial plane is assumed. 
The outer boundaries along each axis are located at 39.6 km.

An atmosphere of small density $\rho = 2\times 10^4~{\rm g/cm^3}$
is added uniformly outside the neutron star and disk at $t=0$, 
since the vacuum is not allowed in grid-based hydrodynamics
implementations. We note that the density of atmosphere can be
chosen to be much smaller than the nuclear density
$\rho_{\rm nuc}$. This is the
advantage of HRC schemes in which such low density can be handled 
in contrast with high-resolution shock-capturing schemes \cite{ShibaFont}.
Since the atmosphere is added as well as a small pressure perturbation
is imposed, the Hamiltonian and momentum constraints are enforced at $t=0$
using the method described in Sec. IV. 

In Fig. \ref{FIG12}(a), we show the evolution of the central density of the
neutron star, and mass and angular momentum of the disk which are
defined by
\beqn
&& M_{\rm disk} \equiv \int_{x \geq x_{\rm in}} \rho_* d^3x,\\
&& J_{\rm disk} \equiv \int_{x \geq x_{\rm in}} S_{\varphi} d^3x,
\eeqn
where $x_{\rm in}$ denotes the initial coordinate radius of the inner
edges of the disk. The figure shows that our implementation 
keeps the equilibrium system to be in equilibrium for more than
$20 P_c$. With the grid of size (241,241), increase of the
density, which is perhaps associated with the outward transport
of the angular momentum, is at most $\sim 1\%$ at $t=20P_c$. 
The change in the baryon rest mass and angular momentum of the disk, 
which is caused spuriously by the mass transfer from the central star and
mass accretion to the central star due to a numerical error, is 
smaller than $\sim 0.1\%$. Also, the numerical results
converge at better than second order with improving the grid resolution. 

In Fig. \ref{FIG12}(b), we also show the evolution of the ADM mass,
angular momentum, and averaged violation of the Hamiltonian
constraint. It is found that the ADM mass is conserved within $\sim 1\%$
error for $t \leq 20P_c$ with $(241, 241)$ grid resolution.
An outstanding feature is that the
angular momentum is conserved with much better accuracy than
the ADM mass. This is a feature when a HRC scheme
is adopted \cite{ShibaFont}. The averaged violation of the
Hamiltonian constraint also remains to be a small magnitude
for $t \leq 20P_c$ and converges at better than second order.
All these results illustrate that our implementation can compute
self-gravitating equilibrium systems accurately. 

\subsection{Winding-up of magnetic field lines
in a disk around a neutron star}

Next, we add magnetic fields confined only in the disk around
the neutron star. For this test, we use the same
system of a neutron star and a disk which is described in Sec. VI B.
Similar test in a fixed background spacetime of a black hole
has been performed in \cite{gammie,VH}. Here, we perform the test
in full general relativity replacing the black hole by a neutron star.
The purpose of this subsection is to illustrate that
our implementation can follow the growth of magnetic fields
by winding-up 
due to differential rotation of the disk. Subsequent papers will focus on
detailed scientific aspect of this issue \cite{SSSS}. 

Following \cite{gammie,VH}, the $\varphi$ component of the
vector potential $A_{\varphi}$ is chosen as
\beqn
A_{\varphi}=\Big\{
\begin{array}{ll}
A (\rho-\rho_{0}) & {\rm for}~\rho \geq \rho_0,\\
0                 & {\rm for}~\rho < \rho_0,
\end{array}
\eeqn
where $A$ is a constant which determines the 
magnetic field strength. Then the magnetic fields are given by 
$\cB^z=x^{-1} \pa_x A_{\varphi}$ and $\cB^x=-x^{-1} \pa_z A_{\varphi}$. 
This choice of $A_{\varphi}$ produces poloidal field loops
that coincide with isodensity contours. Here, $\rho_0$ is chosen
as $0.3\rho_{\rm max:disk}$ where $\rho_{\rm max:disk}$ is the
maximum density inside the disk. In the following, all the simulations
are performed in axial symmetry with (301, 301) grid size and with
the grid spacing of 0.165 km. The reflection symmetry with respect to the
equatorial plane is assumed. We note that the boundary condition for
the magnetic field is $\cB^x=\cB^y=0$ and $\pa_z\cB^z=0$ at the
equatorial plane (in contrast to those for 
velocity fields $v^i$ and $u_i$ for which, e.g., 
$\pa_z v^x=\pa_z v^y=0$ and $v^z=0$ at the equatorial plane). 
Outer boundary conditions are not necessary for the magnetic field
in the present simulations 
since the location of the outer boundary is far enough from the 
center that the magnetic field lines do not reach the outer boundaries. 
The Hamiltonian and momentum constraints are enforced at $t=0$
using the method described in Sec. IV. Since the magnetic field strength
we choose is very weak initially, the obtained initial condition is 
approximately the same as that of no magnetic fields.

\begin{figure}[thb]
\vspace{-4mm}
\begin{center}
\epsfxsize=3.in
\leavevmode
\epsffile{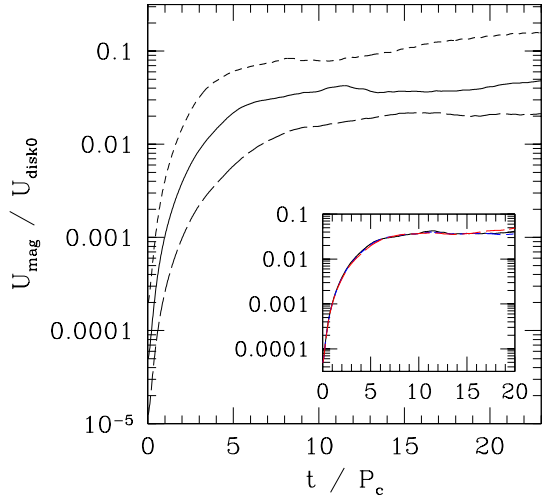} 
\end{center}
\vspace{-4mm}
\caption{
Evolution of $U_{\rm mag}$ in unit of initial internal energy of
$U_{\rm disk0}$ for $R_B=1.3 \times 10^{-5}$ (long-dashed curve), 
$5 \times 10^{-5}$ (solid curve), and 
$2 \times 10^{-4}$ (dashed curve). 
In the smaller panel, we display evolution of $U_{\rm mag}/U_{\rm disk0}$
for $R_B=5 \times 10^{-5}$ with three levels of grid
resolutions with sizes (301, 301) (solid curve),
(241, 241) (dashed curve), and (201, 201) (long-dashed curve).
Three curves approximately agree. 
\label{FIG13} }
\end{figure}

Simulations are performed for various values of $A$ which is chosen
so that the magnetic pressure is initially much smaller than the gas pressure.
In the following we specify the model in terms of the
initial ratio of the energy of magnetic fields to the internal energy
of the disk (hereafter $R_B$) instead of $A$. Here, the energy of magnetic
fields and the internal energy of the disk is simply defined by 
\beqn
&&U_{\rm mag} \equiv \int_{\rm disk} b^2 d^3x,\\ 
&&U_{\rm disk} \equiv \int_{\rm disk} \rho_* \varepsilon d^3x, 
\eeqn
and thus, $R_B \equiv U_{\rm mag}/U_{\rm disk}$ at $t=0$. We note that the
precise definition of the magnetic energy is unknown in general 
relativity, but the present definition is likely to give a guideline
for the magnitude within an error of a factor of $\sim 2$--3. 

In Fig. \ref{FIG13}, we show the evolution of $U_{\rm mag}$ for three
values of $R_B$. Here, the magnetic energy is plotted in units of the
initial value of $U_{\rm disk}$ (hereafter $U_{\rm disk0}$) which is
about $1.8 \times 10^{-4} M_{\rm disk}$. It is found that $U_{\rm
mag}$ grows monotonically until the growth is saturated irrespective
of the value of $R_B$. The growth rate is in proportional to
$R_B^{1/2}$ in the early phase before the saturation is reached. This
indicates that differential rotation winds up the magnetic field lines
for amplifying the field strength \cite{BSS,Stu}.  After the
saturation occurs, $U_{\rm mag}/U_{\rm disk0}$ relaxes to $\sim
0.02$--0.2. These values indicate that the magnetic $\beta$ parameter
often referred in \cite{VH} is of order $\sim 10$. These relaxed values are
in good agreement with previous results obtained in the simulation
with a fixed background \cite{VH}.  The magnetic energy reached after
the saturation depends on $R_B$, indicating that not only the
winding-up of the field lines but also other mechanisms (which may be
MRI or other instabilities associated with the magnetic fields) are likely to
determine the final value.

To check that the growth of the magnetic fields occurs irrespective of
grid resolution, we performed additional simulations for $R_B=5\times
10^{-5}$ with grid sizes of (241, 241) and (201, 201) without changing
the location of the outer boundaries.  In the small panel of
Fig. \ref{FIG13}, evolution of the magnetic energy for these cases as
well as for (301, 301) grid size is displayed. It is shown that the
growth rate depends very weakly on the grid resolution. This confirms
that our simulation can follow the winding-up of the magnetic field
lines well. On the other hand, it should be mentioned that the fastest
growing mode of the MRI cannot be resolved in the present computational
setting since the characteristic
wavelength for this mode $\sim 2\pi v_A/\Omega$, where $v_A$ denotes
the characteristic Alfv\'en speed, is approximately as small as
the grid size (for $R_B=5 \times 10^{-5}$, 
$2\pi v_A/\Omega \approx \Delta x$) 
in the current setting. To resolve the fastest growing mode,
the grid size should be at least one tenth of the present one. 
Performing such a simulation of high resolution
is beyond scope of this paper and an issue for the next step. 

\begin{figure}[t]
\begin{center}
\epsfxsize=2.2in
\leavevmode
\epsffile{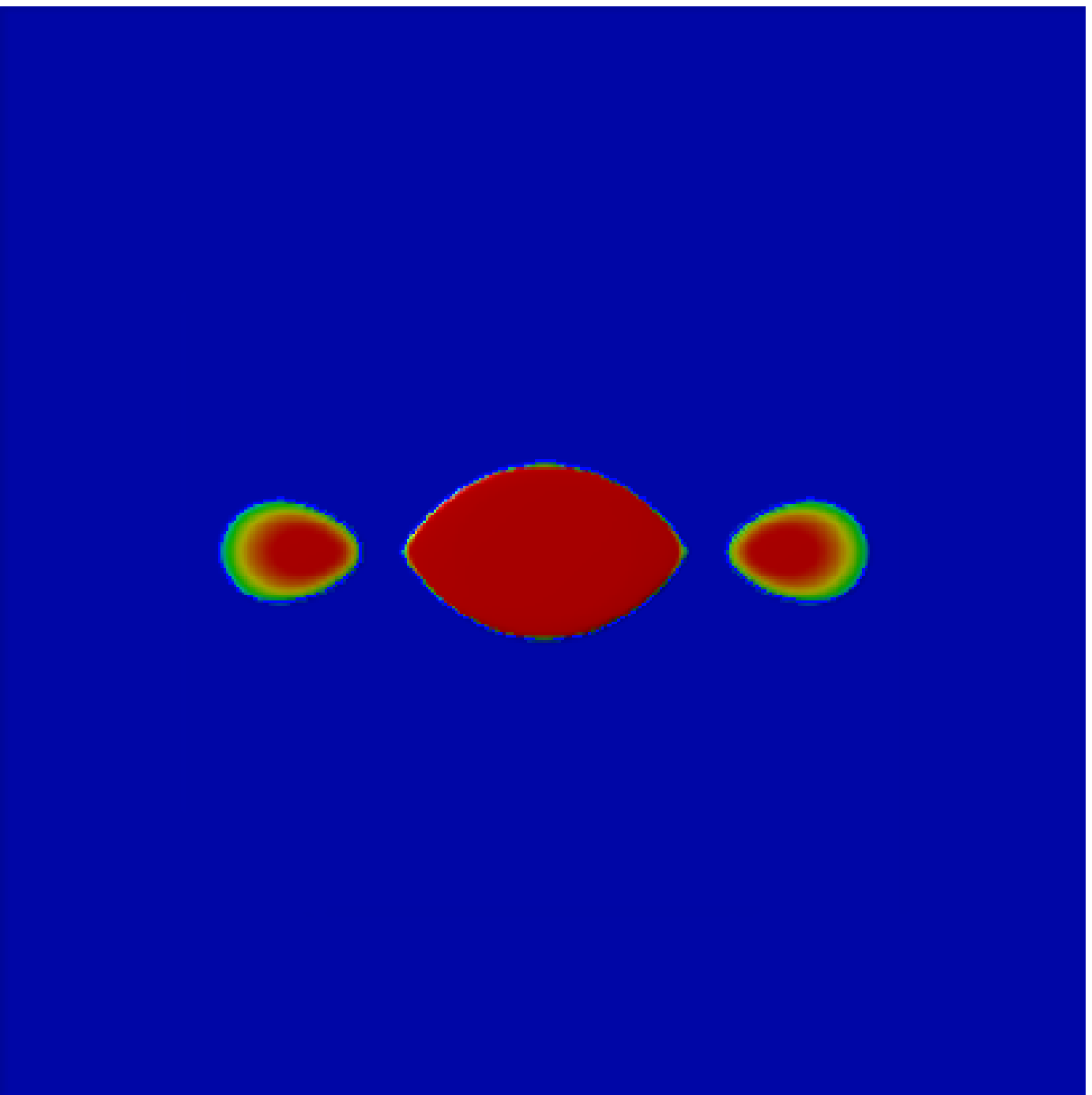} 
\epsfxsize=2.2in
\leavevmode
~\epsffile{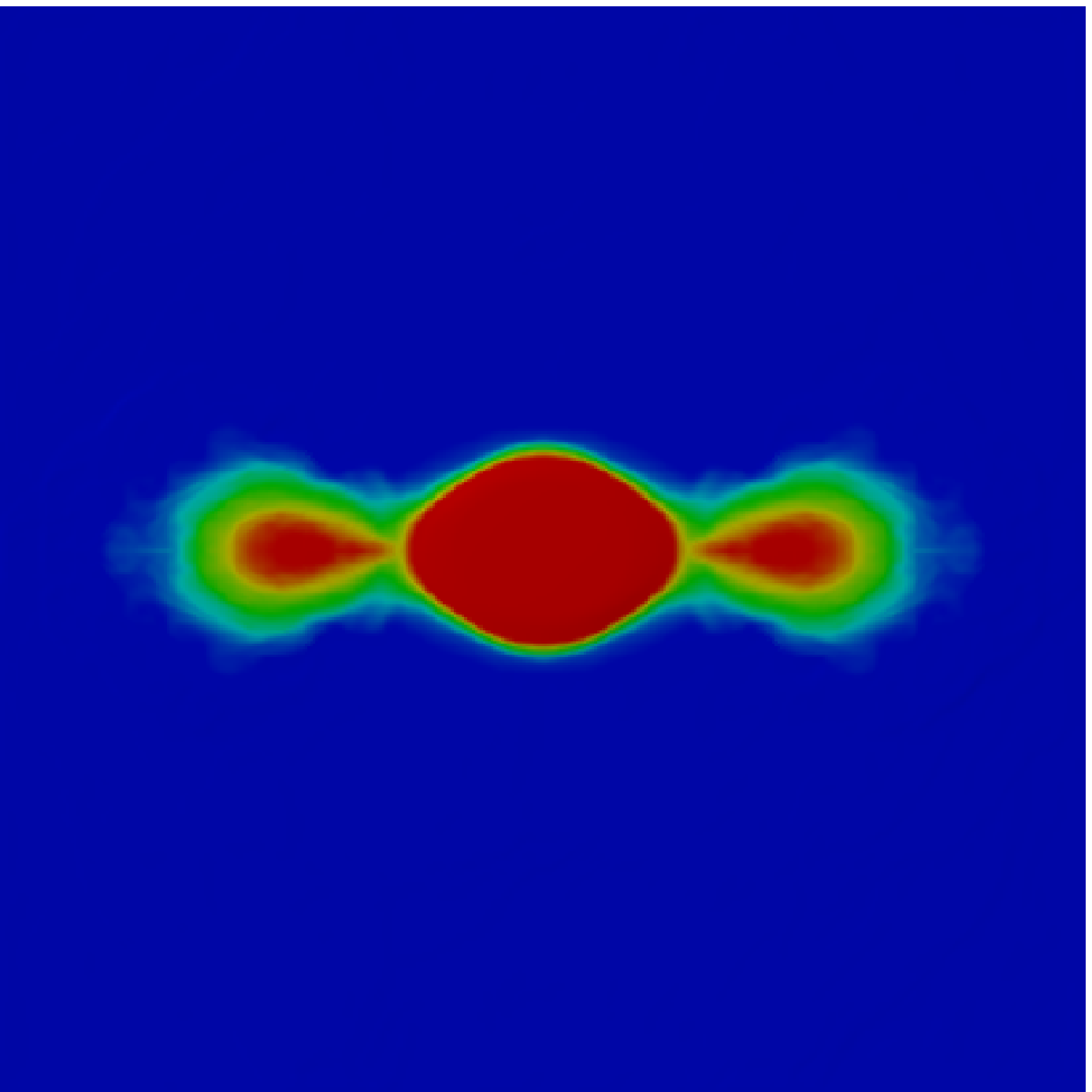} 
\epsfxsize=2.2in
\leavevmode
~\epsffile{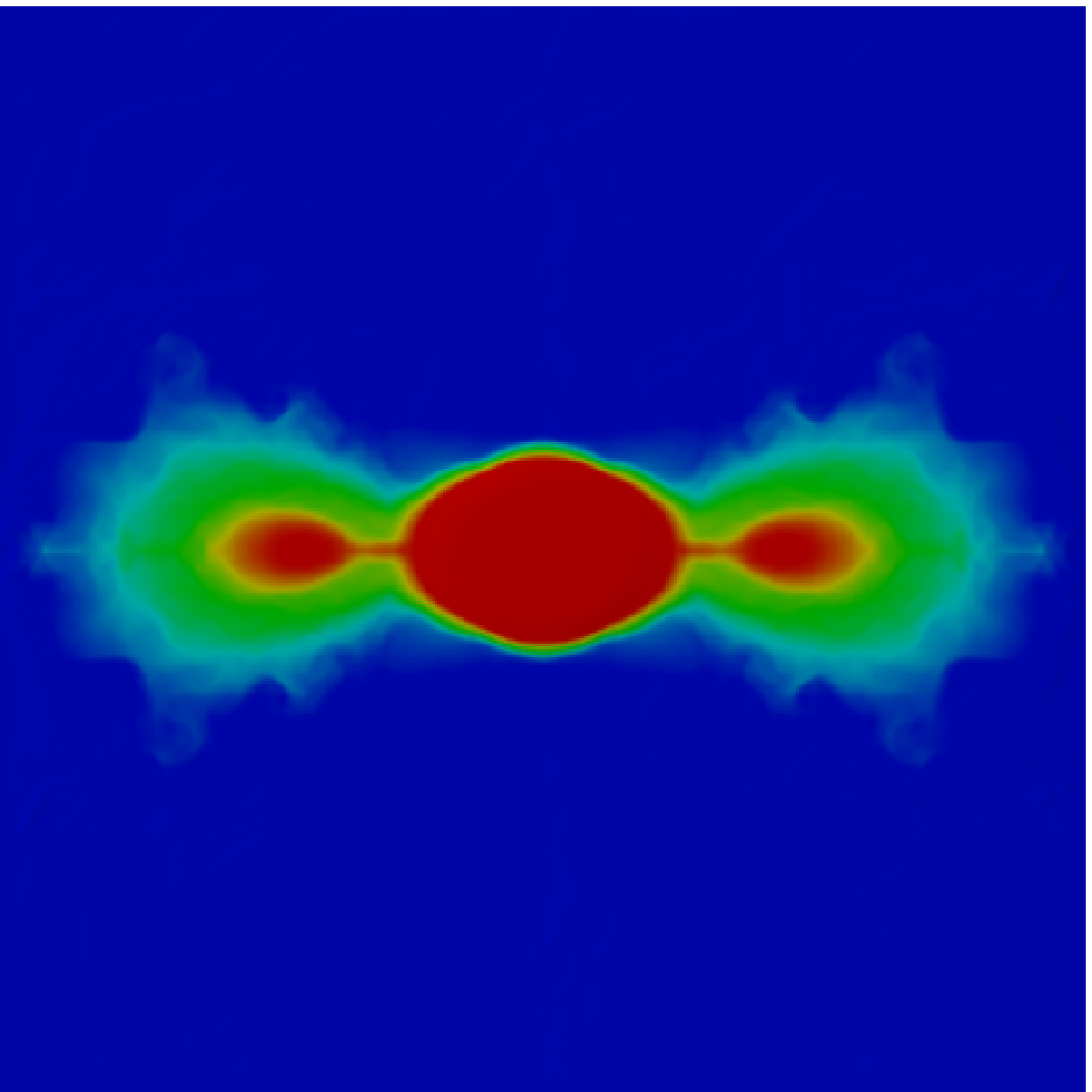} 
\end{center}
\caption{Snapshots of the density profile of a neutron and
a disk in $x-z$ plane at $t=0$, 9.78, and 19.40 ms
for $R_B=2 \times 10^{-4}$. For $\rho > 10^{10}~{\rm g/cm^3}$,
the density is denoted by the same color (red), and
for $10^{10}~{\rm g/cm^3} \geq \rho \geq 10^{7}~{\rm g/cm^3}$,
the color is changed (from red to green). 
\label{FIG14} }
\end{figure}

\begin{figure}[thb]
\vspace{-4mm}
\begin{center}
\epsfxsize=3.in
\leavevmode
\epsffile{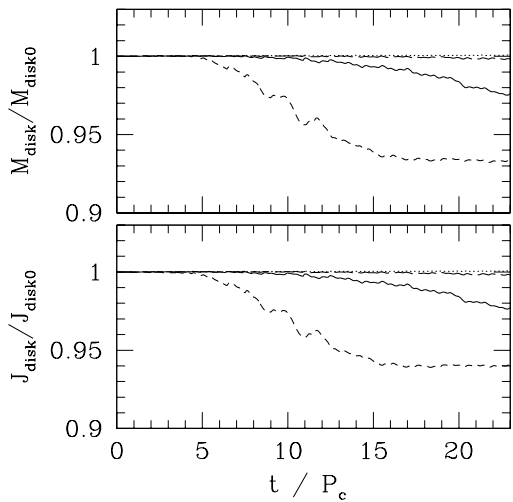} 
\end{center}
\vspace{-6mm}
\caption{Evolution of $M_{\rm disk}$ and $J_{\rm disk}$ for
$R_B=0$ (dotted curves), $1.3\times 10^{-5}$ (long-dashed curves), 
$5\times 10^{-5}$ (solid curves), and 
$2\times 10^{-4}$ (dashed curves).
\label{FIG15} }
\end{figure}

In Fig. \ref{FIG14}, snapshots of the density profile of
disks are displayed for which $R_B=2 \times 10^{-4}$. It shows that 
with the growth of magnetic fields due to winding-up of
the field lines, a wind is induced 
to blow the matter in the outer part of the disk off.
Also, the matter in the inner part of the
disk gradually falls into the neutron star because of the angular momentum
transport by the magnetic fields from the inner to the outer parts
(see Fig. \ref{FIG15}). After the nonlinear development of
the turbulence, the disk settles down to a quasi stationary state.
As explained in \cite{H00,VH}, this is probably due to the imposition of
axial symmetry which precludes the development of the azimuthal unstable
modes. Also, in the present numerical simulation, MRI which could
induce turbulence is not well resolved. This may be also a reason. 

In Fig. \ref{FIG15}, we show the evolution of mass and angular
momentum of the disk. It shows that after the saturation of the
nonlinear growth of the magnetic fields, these quantities
decrease. Decrease rates of the mass and angular momentum take maximum
values soon after the growth of the magnetic pressure is saturated
(e.g., at $t \approx 5 P_c$ for $R_B=2 \times 10^{-4}$; cf. the dashed
curves).  Then, the mass and angular momentum relax to approximately
constants (cf. the dashed curves).  This indicates that the disk
settles down to a quasistationary state. An interesting feature is
that $J_{\rm disk}$ is approximately proportional to $M_{\rm disk}$
throughout the evolution. This is reasonable because the specific
angular momentum $j$ is constant in the disk at $t=0$, and
approximately so is the matter fallen to the neutron star as long as
the magnetic pressure is much smaller than the gas pressure. However,
in the case of $R_B=2 \times 10^{-4}$, at $t \sim 20 P_c$ for which
growth of the magnetic field has already saturated enough, $J_{\rm
disk}/M_{\rm disk}$ slightly deviates from the initial value.  This
indicates that angular momentum is transported by the effect of
magnetic fields.

We also performed a simulation for a toroidal magnetic field $B^{\varphi}$. 
For $B^{\varphi}$, we gave 
\beqn
B^{\varphi}=\Big\{
\begin{array}{ll}
C (\rho-\rho_{0})z/(z+z_0) & {\rm for}~\rho \geq \rho_0,\\
0                 & {\rm for}~\rho < \rho_0,
\end{array}
\eeqn
where $\rho_0=0.3 \rho_{\rm max:disk}$ and 
$z_0$ is a constant much smaller than the scale hight of the disk.
We note that $B^{\varphi}$ has to be zero in the equatorial plane
because we impose a reflection symmetry for the matter field 
with respect to this plane. 
In this simulation, magnetic energy decreases monotonically due to
a small expansion of the disk induced by the magnetic pressure.
In this case, no instability sets in. This is a natural consequence since the
field lines are parallel to the rotational motion, and hence,
they are not wound by the differential rotation. Obviously, 
the assumption of the axial symmetry prohibits deformation of
the magnetic field lines and plays a crucial role for stabilization. 
If a nonaxisymmetric simulation is performed, MRI could 
set in \cite{H00,MHM,HVH}. 

\section{Summary and discussion}

In this paper, we describe our new implementation for ideal GRMHD
simulations. In this implementation, 
Einstein's evolution equations are evolved by a latest version of BSSN
formalism, the MHD equations by a HRC scheme, and the induction
equation by a constraint transport method. We performed a number of
simulations for standard test problems in relativistic MHD including
special relativistic magnetized shocks, general relativistic
magnetized Bondi flow in the stationary spacetime, and fully general
relativistic simulation for a self-gravitating system composed of a
neutron star and a disk. Our implementation yields accurate and
convergent results for all these test problems.  In addition, we
performed simulations for a magnetized accretion disk around a neutron
star in full general relativity. It is shown that magnetic fields in
the differentially rotating disk are wound, and as a result, the
magnetic field strength increases monotonically until a saturation is
achieved. This illustrates that
our implementation can be applied for investigation of growth of
magnetic fields in self-gravitating systems.

In the future, we will perform a wide variety of
simulations including magnetized stellar core
collapse, MRI for self-gravitating
neutron stars and disks, magnetic braking of differentially rotating
neutron stars, and merger of binary magnetized neutron
stars. Currently, we consider that the primary target is stellar core
collapse of a strongly magnetized star to a black hole and a neutron
star which could be a central engine of gamma-ray bursts.  Recently,
simulations aiming at clarifying these high energy phenomena have been
performed \cite{GRB1,GRB2}. In such simulations, however, one
neglects self-gravity and also assumes the configuration of the disks
around the central compact object and magnetic fields without physical
reasons.  On the other hand, Newtonian MHD simulations including
self-gravity consistently have recently performed in \cite{GRB3}. However,
stellar core collapse to a black hole and gamma-ray bursts are
relativistic phenomena. For a self consistent study, it is
obviously necessary to perform a general relativistic simulation from
the onset of stellar core collapse throughout formation of a neutron
star or a black hole with surrounding disks. Subsequent phenomena such
as ejection of jets and onset of MRI of disks should be investigated using the
output of the collapse simulation.  In previous papers \cite{SS,SS05},
we performed fully general relativistic simulations of stellar core
collapse to formation of a neutron star and a black hole in the absence of 
magnetic fields. As an extension of the previous work, simulation for
stellar core collapse with a strongly magnetized massive star should
be a natural next target. 

It is also important and interesting to clarify how MRI sets in and 
how long the time scale for the angular momentum
transport after the onset of the MRI
is in differentially rotating neutron stars. Recent 
numerical simulations for merger of binary neutron stars in full
general relativity \cite{STU,STU2} have clarified that if total mass
of the system is smaller than a critical value, the outcome after the
merger will be a hypermassive neutron star for which the self-gravity
is supported by strong centrifugal force generated by rapid and
differential rotation.  Furthermore, the latest simulations have
clarified that the hypermassive neutron star is likely to have an
ellipsoidal shape with a large ellipticity \cite{STU2}, implying that
it can be a strong emitter of high-frequency gravitational waves which
may be detected by advanced laser interferometric gravitational wave
detectors \cite{S05}. In our estimation of amplitude of gravitational
waves \cite{S05}, we assume that there is no magnetic field in the
neutron stars. However, the neutron stars in nature are magnetized,
and hence, the hypermassive neutron stars should also be.  If the
differential rotation of the hypermassive neutron stars amplifies the
seed magnetic field via winding-up of magnetic fields or MRI very
rapidly, the angular momentum may be redistributed and hence the
structure of the hypermassive neutron stars may be significantly
changed.  In \cite{STU2}, we evaluate the emission time scale of
gravitational waves for the hypermassive neutron stars is typically
$\sim 50$--$100$ ms for the mass $M \sim 2.4$--$2.7 M_{\odot}$
assuming the absence of the magnetic effects. Here, the time scale of
$\sim 50$--$100$ ms is an approximate dissipation time scale of
angular momentum via gravitational radiation, and hence in this case,
after $\sim 50$--$100$ ms, the hypermassive neutron stars collapse to
a black hole because the centrifugal force is weaken. Thus, it is
interesting to ask if the dissipation and/or transport time scale of
angular momentum by magnetic fields is shorter than $\sim 50$--$100$
ms so that they can turn on before collapsing to a black
hole. Rotational periods of the hypermassive neutron stars are 0.5--1
ms. Thus, if the magnetic fields grow in the dynamical time scale
associated with the rotational motion via MRI, the amplitude and
frequency of gravitational waves may be significantly
affected. According to a theory of MRI \cite{BaHa}, the wavelength of
the fastest growing mode is $\sim 10 (B/10^{12}~{\rm gauss})
(\rho/10^{15}~{\rm g/cm^3})^{-1/2} (P/1~{\rm ms})$ cm where $B$,
$\rho$, and $P$ denotes a typical magnetic field strength, density,
and rotational period, respectively. This indicates that a turbulence
composed of small eddies (for which the typical scale is
much smaller than the stellar radius) will set in. 
Subsequently, it will contribute to a secular angular momentum
transport for which the time scale is likely to be longer than the
growth time scale of MRI $\sim $ a few ms although it is not clear if
it is longer than $\sim 100$ ms. On the other hand, if the transport 
time scale is not as short as $\sim 100$ ms, 
other effects associated with magnetic fields will not affect the
evolution of the hypermassive neutron stars. Indeed, Ref. \cite{BSS}
indicates that the typical time scale associated with magnetic braking 
(winding-up of magnetic field lines) depends on the initial strength 
of the magnetic fields, and it is much longer than the dynamical time 
scale as $\sim 100~(10^{12}~{\rm gauss}/B)$ s. In this case, the 
hypermassive neutron stars can be strong emitters of gravitational 
waves as indicated in \cite{S05}.  As is clear from this discussion, 
it is important to clarify the growth time scale of magnetic fields in
differentially rotating neutron stars. This is also the subject in our
subsequent papers \cite{SSSS}.

\acknowledgments

We are grateful to Stu Shapiro for many valuable discussions and  
to Yuk-Tung Liu for providing solutions for Alfv\'en
wave tests presented in Sec. V and valuable discussions. 
We also thank Miguel Aloy, Matt Duez, Toni Font, S. Inutsuka, 
A. Mizuta, Branson Stephens, and R. Takahashi 
for helpful discussions. Numerical computations were 
performed on the FACOM VPP5000 machines at the data processing center
of NAOJ and on the NEC SX6 machine in the data processing center of
ISAS in JAXA.  This work was in part supported by Monbukagakusho Grant
(Nos. 15037204, 15740142, 16029202, 17030004, and 17540232).

\end{document}